\documentclass[aps,prc,twocolumn,showpacs,preprintnumbers,
               nofootinbib,float,superscriptaddress,longbibliography]{revtex4-1}
\usepackage{graphicx, fancybox}
\usepackage{amsmath,amssymb}
\usepackage[colorlinks=true, pdfstartview=FitV, linkcolor=red, citecolor=blue, urlcolor=blue]{hyperref}
\usepackage{color}
\usepackage{soul}
\usepackage{url}
\usepackage[utf8]{inputenc}

\usepackage[normalem]{ulem}

\begin{document}

\title{Multi-particle and charge-dependent azimuthal correlations in heavy-ion collisions\\ at the Relativistic Heavy-Ion Collider}

\author{Bj\"orn Schenke}
\affiliation{Physics Department, Brookhaven National Laboratory, Upton, NY 11973, USA}

\author{Chun Shen}
\affiliation{Department of Physics and Astronomy, Wayne State University, Detroit, Michigan, USA}
\affiliation{RIKEN BNL Research Center, Brookhaven National Laboratory, Upton, NY 11973, USA}

\author{Prithwish Tribedy}
\affiliation{Physics Department, Brookhaven National Laboratory, Upton, NY 11973, USA}

\begin{abstract}
  We study multi-particle azimuthal correlations in relativistic heavy-ion collisions at a center of mass energy of 200 GeV. We use the IP-Glasma model to initialize the viscous hydrodynamic simulation MUSIC and employ the UrQMD transport model for the low temperature region of the collisions. In addition, we study effects of local charge and global momentum conservation among the sampled particles. With the exception of the lowest order three particle correlator $C_{112}$, our framework provides a good description of the existing charge inclusive azimuthal correlation data for Au+Au and U+U collisions at RHIC. We also present results for charge dependent two and three particle correlators in Au+Au and U+U collisions, and make predictions for isobar (Ru+Ru $\&$ Zr+Zr) collisions to provide a much needed baseline for the search for the Chiral Magnetic Effect at RHIC.
\end{abstract}

{\maketitle}

\section{Introduction}

Hydrodynamic modeling has been very successful in describing the bulk properties of the matter formed in relativistic heavy ion collisions. Apart from the centrality and system dependence of particle numbers and average transverse momenta, the centrality and transverse momentum dependence of the experimental data on $n$'th order harmonic flow coefficients $v_{n}$ have been well reproduced by hydrodynamic simulations. Such studies have provided important insights into initial state geometry, fluctuations, and the transport parameters such as shear and bulk viscosity of the medium created in relativistic heavy ion collisions. It must be noted that many of such data-model comparisons have been only restricted to $v_{n}$ extracted from charge inclusive azimuthal angular correlations, typically involving only two particles. 

Recent experimental results from RHIC (and the LHC) go beyond the conventional two-particle correlations and focus on the precision measurements of multi-particle azimuthal correlations, including their charge dependence~\cite{STAR:2018fpo,Adamczyk:2017hdl,Adamczyk:2017byf,Aad:2014fla,ALICE:2016kpq}. 

One of the primary motivations for measuring multi-particle correlations is to reduce the 
relative contribution of non-flow correlations~\cite{Ollitrault:1997vz,Poskanzer:1998yz} in the measurements of harmonic flow coefficients $v_{n}$~\cite{Borghini:2000sa,Borghini:2001vi}. Non-flow correlations are all correlations emerging from effects other than the collective motion of the fluid, including intrinsic quantum correlations, correlations from Coulomb interactions, resonance decays, local charge and momentum conservation during particle formation, etc., all of which scale inversely with multiplicity $N$ to some power.

One therefore expects harmonic flow coefficients $v_{n}\{ m > 2 \}$ extracted using $m$-particle correlations to have less non-flow contributions as compared to conventional two-particle $v_{n}\{2\}$ coefficients~\cite{Borghini:2000sa}. Consequently, observables based on multi-particle azimuthal correlations are better suited for comparison to purely hydrodynamic simulations of heavy ion collisions. 
   
Apart from the reduction of non-flow correlations, certain multi-particle azimuthal correlators can provide additional information compared to the conventional two- and multi-particle harmonic flow observables. In particular, correlators that involve harmonics of different order can provide a measure of correlations between the flow vectors of different harmonic order. Such correlators have been argued to be able to put stronger constraints on initial state models, to separate linear from non-linear hydrodynamic response, and to provide more information on the temperature dependence of $(\eta/s) (T)$ \cite{Teaney:2010vd, Qiu:2012uy, Jia:2012ma, Teaney:2013dta, Bhalerao:2013ina,Niemi:2015qia,Giacalone:2016afq,Gardim:2016nrr}. The study of three and four particle correlations within our hybrid framework, comprised of IP-Glasma initial state, viscous hydrodynamics, and microscopic transport, is one primary objective of this work.

Other quantities of experimental interest that have been extensively studied at the RHIC and LHC are charge dependent azimuthal correlations~\cite{Abelev:2009ac,Abelev:2009ad,Abelev:2012pa,Adamczyk:2013hsi, Adamczyk:2014mzf, Adam:2015vje, Khachatryan:2016got, Acharya:2017fau,Sirunyan:2017quh}. In particular, when including charge dependence, certain multi-particle correlators are sensitive to signals of the chiral magnetic effect (CME) \cite{Kharzeev:1998kz,Kharzeev:2004ey, Kharzeev:2007jp,Voloshin:2004vk, Fukushima:2008xe, Ajitanand:2010rc}. %\bps{maybe write a bit on other explicit calculations}. 
However, it has been demonstrated that such observables are also sensitive to background effects, i.e., non-CME phenomena that lead to charge-dependent azimuthal correlations~\cite{Voloshin:2004vk,Pratt:2010gy, Pratt:2010zn, Schlichting:2010qia}. It is thus of the utmost importance to have a clear expectation for such background correlations in a framework where no CME signal is present~\cite{Pratt:2010gy, Pratt:2010zn, Schlichting:2010qia, Ma:2011uma, Deng:2018dut}. This work presents results for such a baseline for CME observables for heavy ion collisions at top RHIC energy, including predictions for forthcoming isobar system collisions Ru+Ru and Zr+Zr~\cite{Voloshin:2010ut,Skokov:2016yrj}.

The paper is organized as follows. We present the hybrid model used for our calculations in Section \ref{sec:model}, in which we briefly discuss the initial state IP-Glasma, matching to hydrodynamics, the viscous hydrodynamic simulation \textsc{Music}, as well as matching to the microscopic hadronic transport model UrQMD. In Section \ref{sec:incl} we focus on charge inclusive and identified particle observables, presenting observables such as multiplicity and mean transverse momentum distributions with centrality, as well as anisotropic flow harmonics from two and four particle correlations, which were used to constrain all model parameters. We then present results for mixed harmonic correlators, including symmetric cumulants. In Section \ref{sec:charge} we present charge dependent multi-particle correlators, including those relevant for CME searches. We also present predictions for isobar systems in this section. We conclude in Section \ref{sec:conc}.

\section{Framework}\label{sec:model}

The initial state and early time dynamics in a system governed by quantum chromodynamics (QCD) in the high energy limit is unambiguously (to leading order) described by a classical system of gluons obeying the Yang Mills equations \cite{Gelis:2010nm}. A numerical evaluation of the Yang Mills equations, given the color currents of the two incoming nuclei, allows to extract the leading order result for the produced gluon fields a short time (fractions of 1\,{\rm fm}/c) after the collision without further approximations. Augmented by a transverse geometry and energy dependence constrained by deep inelastic scattering data, the impact parameter dependent Glasma model (in short IP-Glasma) \cite{Schenke:2012wb,Schenke:2012hg} evaluates the Yang-Mills equations and generates the boost invariant energy momentum tensor of gluon fields $T^{\mu\nu}_{\rm CYM}$ to be used as input in hydrodynamic simulations at time $\tau_0$.

We note that in order to generate the color charge density for the incoming nuclei, nucleons are sampled from a Woods-Saxon distribution \cite{Woods:1954zz}
\begin{equation}
    \rho(r,\theta) = \frac{\rho_0}{1+\exp[(r-R'(\theta))/a]},
\end{equation}
where nuclear deformation is introduced by
\begin{equation}
    R'(\theta) = R[1+\beta_2 Y_2^0(\theta) +\beta_4 Y^0_4(\theta)].
\end{equation}
Here $\rho_0$ is the nucleon density at the center of the nucleus, $a$ characterizes the diffuseness of the nuclear surface, and $R$ is the (angle averaged) nuclear radius. In the case of the Au, U, and Ru nuclei, we include a deformation, quantified by the parameters $\beta_l$ ($l=2,4$), which multiply the spherical harmonic functions $Y_l^0(\theta)$. All relevant parameters are listed in Table \ref{tab:WSparam}.
\begin{table}[hb]
\begin{center}  
    \begin{tabular}{ | c | c | c | c | c |}
    \hline  
    Nucleus & $R$ [fm] & $a$ [fm] & $\beta_2$ & $\beta_4$ \\ \hline
    $^{238}$U  & 6.81  & 0.55  & 0.28  & 0.093 \\ \hline
    $^{197}$Au & 6.37  & 0.546 & -0.13 & -0.03 \\ \hline
    $^{96}$Ru & 5.085 & 0.46  & 0.158 & 0     \\ \hline 
    $^{96}$Zr & 5.02  & 0.46  & 0     & 0     \\ 
    \hline
    \end{tabular}
\end{center}
\caption{Parameter values used in the Woods-Saxon parametrizations of the five nuclei studied \cite{Filip:2009zz,Masui:2009qk,Hirano:2012kj,Shen:2014vra,Schenke:2014tga,Pritychenko:2013gwa,Goldschmidt:2015qya,Goldschmidt:2015kpa}. \label{tab:WSparam}}
\end{table}
We note that from density functional theory calculations, the nuclear density distributions of Zr and Ru can exhibit larger differences than in our prescription \cite{Xu:2017zcn}. 

In this work we follow the procedure described in \cite{Mantysaari:2017cni}, where for the first time the full energy momentum tensor, including shear stress contributions, was included along with subnucleonic fluctuations.
The initial shear stress tensor, which, as the energy density and flow velocity, is a function of the transverse coordinate, is given by
\begin{equation}
\pi^{\mu\nu}=T^{\mu\nu}_{\rm CYM}-\frac{4}{3}\varepsilon u^\mu u^\nu + \frac{\varepsilon}{3}g^{\mu\nu}\,,
\end{equation}
where the initial energy density $\varepsilon$ and flow velocity $u^\mu$ is obtained by solving the eigenvalue problem $u_\mu T^{\mu\nu}_{\rm CYM}=\varepsilon u^\nu$. We further include a correction in the initial value for the bulk component of the stress energy tensor to account for the difference between the equation of state in the Yang-Mills system ($\varepsilon=3P$, with $P$ the pressure) and in the hydrodynamic simulation, where it is constructed from lattice QCD data and a hadron resonance gas model. This initial value for the bulk pressure $\Pi$, which is diminished within one bulk relaxation time, leads to an additional outward push. We comment in Appendix \ref{app:A} on the quantitative effect of this method of precise matching the energy momentum tensor.

Starting from time $\tau_0$, which in this work we choose to be $0.4\,{\rm fm}$, the energy momentum tensor is evolved according to the conservation law,
\begin{equation}
    \partial_\mu T^{\mu \nu} = 0.
\end{equation}
The viscous parts are evolved with
\begin{equation}
    \tau_\Pi \dot{\Pi} + \Pi = -\zeta \theta - \delta_{\Pi\Pi} \Pi \theta + \lambda_{\Pi\pi} \pi^{\mu\nu} \sigma_{\mu\nu}
\end{equation}
\begin{eqnarray}
    \tau_\pi \Delta^{\mu\nu}_{\alpha \beta} \dot{\pi}^{\alpha \beta} + \pi^{\mu\nu} &=& 2 \eta \sigma^{\mu\nu} - \delta_{\pi\pi} \pi \theta  + \phi_7 \pi^{\langle \mu}\,_\alpha \pi^{\nu \rangle \alpha} \notag \\
     && - \tau_{\pi\pi }\pi^{\langle \mu}\,_\alpha \sigma^{\nu \rangle \alpha} + \lambda_{\pi\Pi} \Pi \sigma^{\mu\nu}.
\end{eqnarray}

The specific implementation to solve these hydrodynamic equations is the simulation \textsc{Music} \cite{Schenke:2010nt,Schenke:2010rr,Schenke:2011bn}, which employs a Kurganov-Tadmor algorithm \cite{Kurganov:2000}.

%=======================================
\begin{figure}[ht!]
  \centering
  \includegraphics[width=0.95\linewidth]{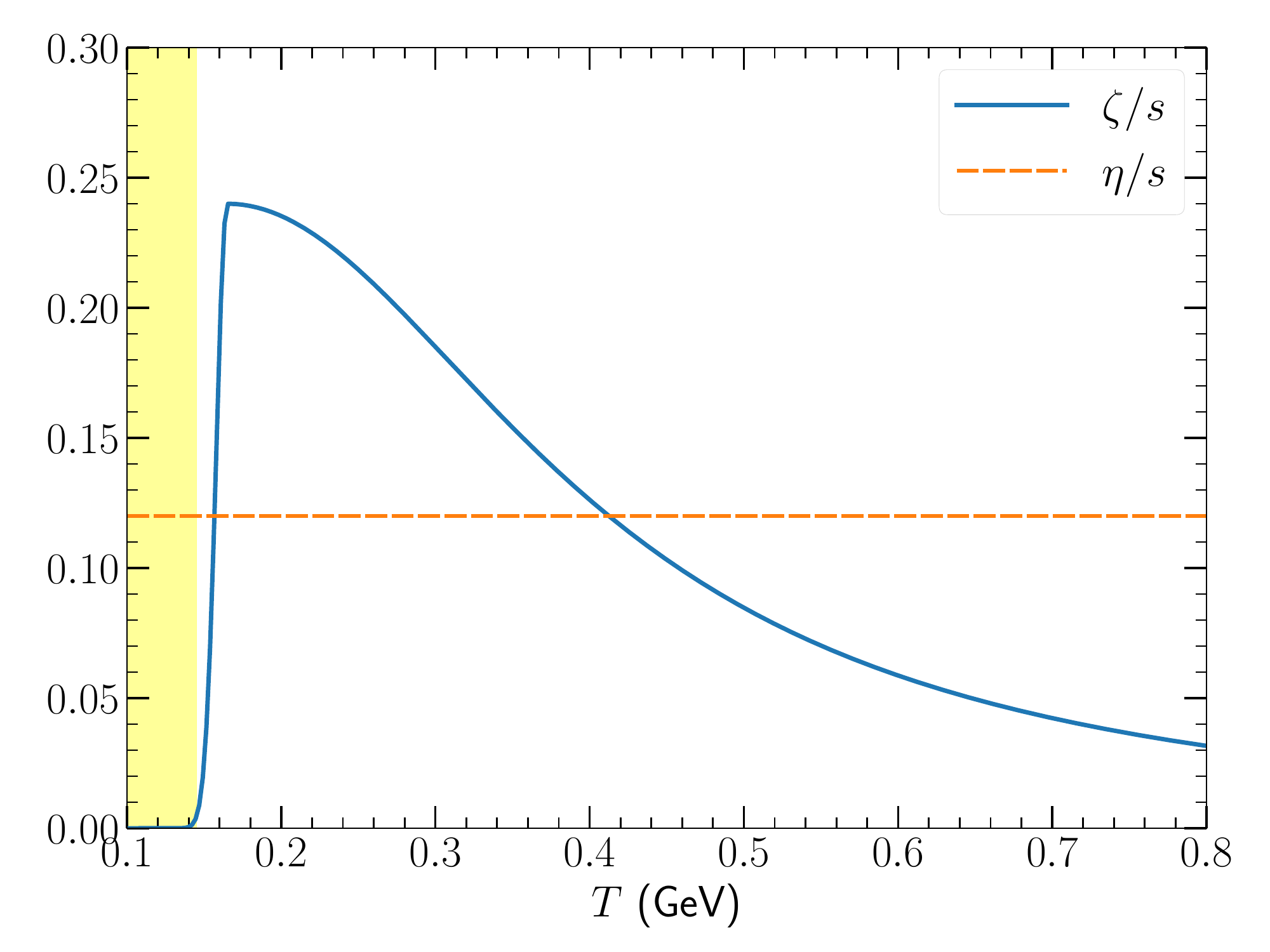}
  \caption{The temperature dependence of the specific shear and bulk viscosity used in the hydrodynamic simulations. The yellow region indicates $T < 145$ MeV, where we employ a hadronic transport model.}
  \label{fig1}
\end{figure}
%=======================================
%
The temperature dependent specific shear and bulk viscosity of the quark gluon plasma (QGP) are chosen as in Fig.~\ref{fig1}. A constant effective $\eta/s = 0.12$ was fixed by matching to the measured anisotropic flow coefficients of charged hadrons. A temperature dependent specific bulk viscosity was introduced to reproduce the mean $p_T$ measurements for identified particles (see discussion in the next section). We use a parametrization in the following form,
\begin{equation}
    \frac{\zeta}{s}(T) = \left\{
        \begin{array}{rcl}
            B_\mathrm{norm}\frac{B_\mathrm{width}^2}{(T/T_\mathrm{peak} - 1)^2 + B_\mathrm{width}^2} & \mbox{ for } & T > T_\mathrm{peak} \\
         B_\mathrm{norm} \exp\left[-\left(\frac{T - T_\mathrm{peak}}{T_\mathrm{width}}\right)^2 \right] & \mbox{ for } & T < T_\mathrm{peak} 
        \end{array} \right. .
\end{equation}
Here the peak of $\zeta/s$ is chosen at $T_\mathrm{peak} = 165$ MeV with a maximum value of $B_\mathrm{norm} = 0.24$. The width of $\zeta/s$ is controlled by the two parameters $B_\mathrm{width} = 1.5$ and $T_\mathrm{width} = 50$ MeV. 
The yellow band in the figure indicates the temperature region ($T\lesssim 145\,{\rm MeV}$) where we no longer employ hydrodynamics but evolve individual hadrons microscopically. 

We use a modern QCD equation of state (EoS) based on continuum extrapolated lattice calculations at zero net baryon chemical potential published by the HotQCD Collaboration \cite{Bazavov:2014pvz}. It is smoothly matched to a hadron resonance gas EoS in the temperature region between 110 and 130 MeV \cite{Moreland:2015dvc}.

Compared to the values of the transport coefficients used in Ref.~\cite{Ryu:2015vwa,McDonald:2016vlt,Ryu:2017qzn,McDonald:2017ayb}, the specific shear viscosity used in this work is approximately 30\% larger. This is mainly because we use a different EoS in this work. The new EoS has a larger speed of sound in the transition region than the previously used s95p-v1 \cite{Huovinen:2009yb}. Consequently, a larger $\eta/s$ is needed to suppress the anisotropic flow during the hydrodynamic evolution. The previously discussed inclusion of an initial effective bulk pressure, that accounts for the difference in pressure between the initial state and hydrodynamic side of the matching, also increases the flow, which must be compensated by larger viscosities.
Another difference to the transport parameters used in \cite{Ryu:2015vwa} is the peak position of the specific bulk viscosity, which was shifted to $T = 165$ MeV, and the width of the peak in $\zeta/s(T)$, which is approximately 5 times wider compared to the parametrization used in \cite{Ryu:2015vwa}. Such a wide peak of $\zeta/s(T)$ is preferred to reproduce the centrality dependence of the measured mean $p_T$, especially in very peripheral collisions.

To ensure enough numerical accuracy, we performed the hydrodynamic simulations on a lattice  with $dx = dy = 0.067$ fm and $d\tau = 0.005$ fm/c.

To describe the dilute hadronic phase, individual fluid cells are converted to hadrons at a switching energy density $\varepsilon_\mathrm{sw} = 0.18$ GeV/fm$^3$. It corresponds to a local temperature of approximately 145 MeV. The produced hadrons are then fed into UrQMD \cite{Bass:1998ca,Bleicher:1999xi}, which simulates hadronic scatterings and decays. Every hydrodynamic switching hyper-surface is sampled multiple times to increase statistics. The number of oversampling events for every hydrodynamic event is determined to ensure at least 100,000 particles are sampled within one unit of rapidity. 

For every particle sample, we can impose global momentum conservation (GMC) and local charge conservation (LCC) at the microscopic level. To do so, we first sample particles independently. After all the particles are generated, we compute the net momentum of the system, $\langle \vec{p} \rangle$. Then we correct every particle's momentum by
\begin{equation}
    \vec{p}_j\,^\prime = \vec{p}_j - w_j \langle \vec{p} \rangle.
\end{equation}
We choose a weight $w_j = p_{T, j}^2/\langle p_T^2 \rangle$ for every particle $j$ in the sample, which is independent of the particle's momentum rapidity. With this choice of weight, the shifts in a particle's transverse momentum and azimuthal angle are
\begin{eqnarray}
    \frac{\delta p_{j, T}}{p_{j,T}} &\approx& - \frac{p_{j,x} \langle p_x \rangle + p_{j, y} \langle p_y \rangle}{\langle p_T^2 \rangle}\\
    \delta \phi_j &\approx& \frac{p_{j,x} \langle p_y \rangle - p_{j, y} \langle p_x \rangle}{\langle p_T^2 \rangle}\,,
\end{eqnarray}
neglecting terms of order $\langle p_{x/y} \rangle^2$.
The correction to every particle's momentum is on the order of $1/N$ where $N$ is the total number of particles in the sampled event.

For the local charge conservation (LCC), we follow the numerical implementation first proposed by Bozek and Broniowski in Ref.~\cite{Bozek:2012en}. In this simple model, charged hadron-antihadron pairs are chosen to be produced at the same space-time point (zero correlation length) with their momenta sampled independently in the local rest frame of the fluid cell. This procedure will maximize the correlations between opposite sign pairs, and more sophisticated prescriptions which incorporate a finite correlation length could weaken the effect of LCC in the studied observables.

These two effects are implemented in the open-source particle sampler package, \texttt{iSS}\footnote{The code package can be downloaded from \url{https://github.com/chunshen1987/iSS}.} \cite{Shen:2014vra,Denicol:2018wdp}.

In the following, many observables will be presented by scaling with the number of participants $N_\mathrm{part}$. The values of $N_\mathrm{part}$ in different centrality bins are determined using a Monte-Carlo Glauber model and listed in Table.~\ref{tab:Npart} for the four studied collision systems.

\begin{table}[h!]
\begin{center}  
    \begin{tabular}{ | c | c | c | c | c |}
    \hline  
    Centrality & Au+Au & U+U & Ru+Ru & Zr+Zr \\ \hline
    0-5\%  & 350.6  & 416  & 165.7  & 165.7 \\ \hline
    5-10\% & 298.6  & 358 & 145.1 & 145.1 \\ \hline
    10-20\% & 234.3 & 280  & 116 & 116     \\ \hline 
    20-30\% & 167.6  & 197  & 84.5 & 84.5     \\ \hline
    30-40\% & 117.1  & 134  & 60 & 60     \\ \hline
    40-50\% & 78.3  & 87  & 41.1 & 41.1     \\ \hline
    50-60\% & 49.3  & 53  & 27 & 27     \\ \hline
    60-70\% & 28.8  & 29  & 16.7 & 16.7     \\ \hline
    70-80\% & 15.7  & 15  & 10 & 10     \\ \hline
    80-90\% & 7  & 7 & 5.8 & 5.8     \\ \hline
    \end{tabular}
\end{center}
\caption{Number of participants as a function of collision centrality for Au+Au, U+U, Ru+Ru, and Zr+Zr collisions from a Monte-Carlo Glauber model. \label{tab:Npart}}
\end{table}

\section{Charge inclusive observables} \label{sec:incl}

In this section, we present results from our simulations and compare with experimental measurements from RHIC. We further make predictions for observables in the Ru+Ru and Zr+Zr runs performed at RHIC in 2018.

\subsection{Particle yields and flow observables}

%
%=======================================
\begin{figure}[ht!]
  \centering
   \includegraphics[width=0.95\linewidth]{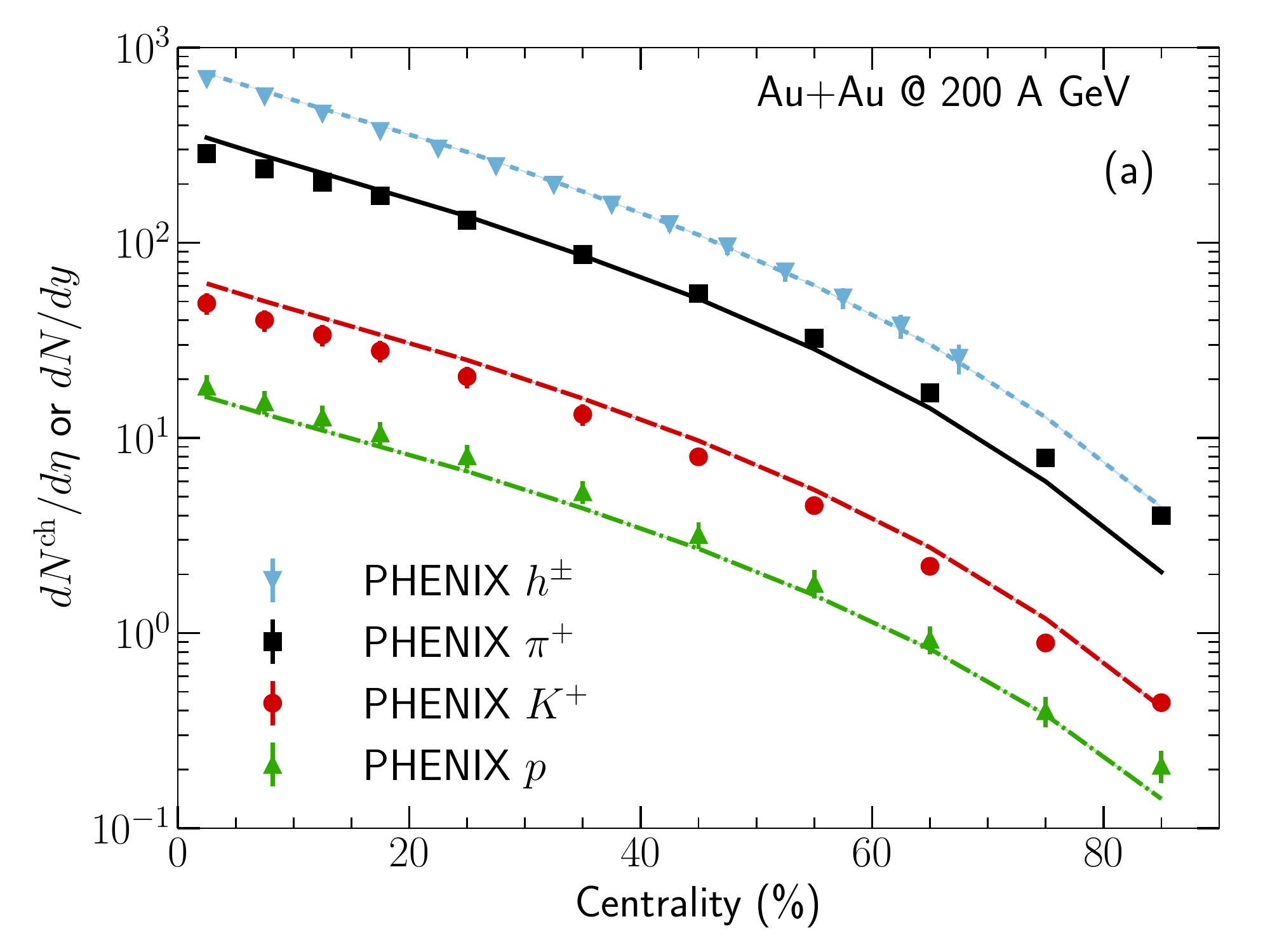} 
   \includegraphics[width=0.95\linewidth]{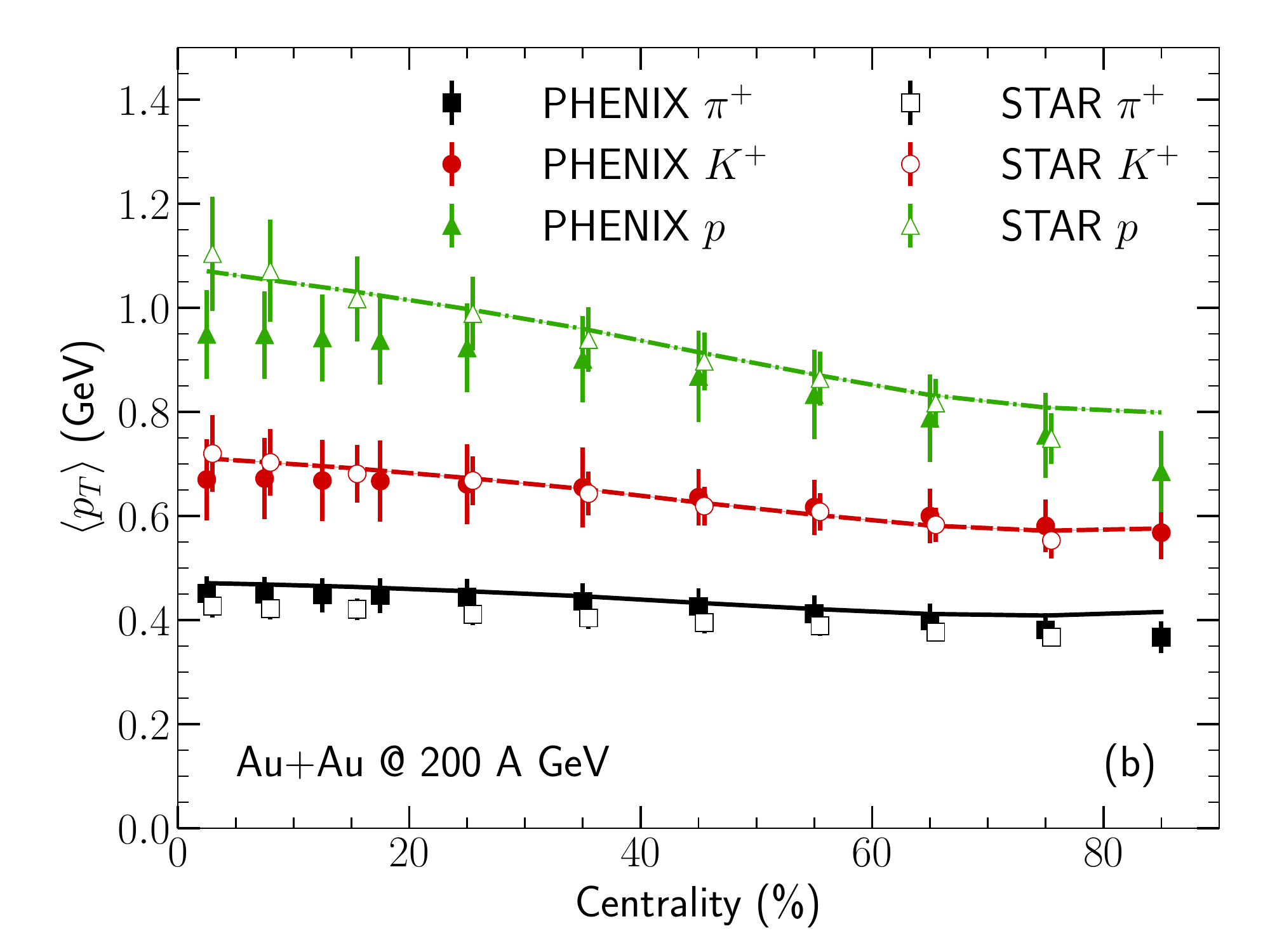}
  \caption{Charged hadron and identified particle yields (a) and their averaged transverse momentum $\langle p_T \rangle$ (b) compared with the PHENIX and STAR measurements \cite{Adler:2003cb,Abelev:2008ab} as a function of centrality in Au+Au collisions at 200 GeV.}
  \label{fig2}
\end{figure}
%=======================================
%
Our theory calculation is calibrated by the produced particle yields and the average transverse momentum, which are shown in comparison to experimental data from the PHENIX \cite{Adler:2003cb} and STAR \cite{Abelev:2008ab} Collaborations in Fig.~\ref{fig2}. In Fig.~\ref{fig2}a, an overall normalization factor on the system's energy-momentum tensor was determined by fitting the charged hadron multiplicity $dN^\mathrm{ch}/d\eta$ in the 0-5\% centrality bin. This freedom of normalization results from the fact that the value of the strong coupling constant is not determined from first principles in the IP-Glasma framework. The remaining centrality dependence of the particle yields is predicted by the IP-Glasma model. The switching energy density $e_\mathrm{sw} = 0.18$ GeV/fm$^3$ ($T_\mathrm{sw}\approx145$ MeV), was adjusted to reproduce experimentally measured proton yields. 

The temperature dependence of the bulk viscosity $\zeta/s(T)$ in Fig.~\ref{fig1} was determined to reproduce the centrality dependence of the identified particle mean $p_T$ shown in Fig.~\ref{fig2}b. We found that a parametrization of $\zeta/s(T)$ with a narrow peak leads to an increase of particle mean $p_T$ when the collision centrality is larger than 70\%, and thus are using a broader peak in the temperature dependence compared to \cite{Ryu:2015vwa}.

\begin{figure}[ht!]
  \centering
  \includegraphics[width=0.95\linewidth]{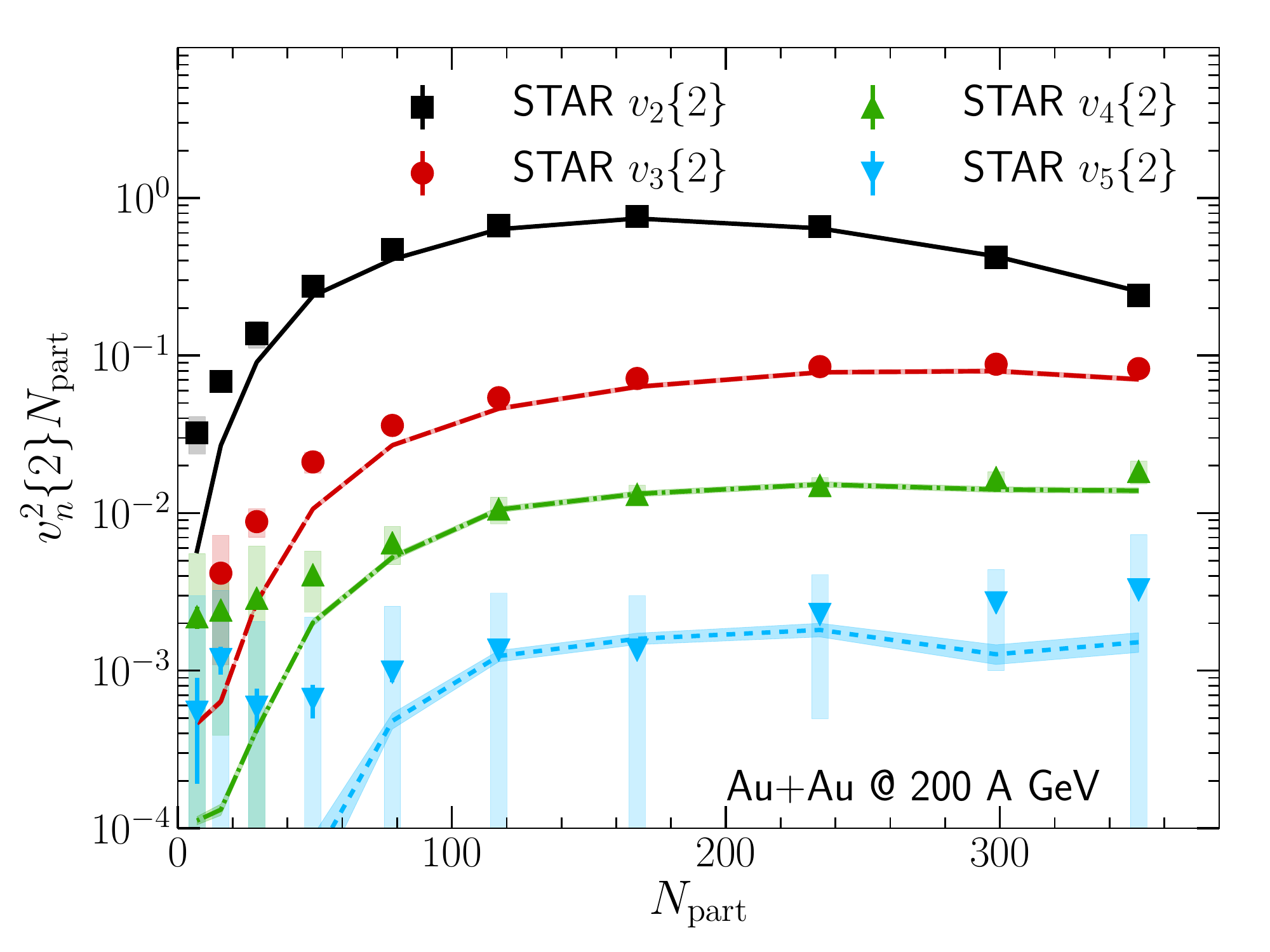}
  \caption{The two-particle anisotropic flow coefficient $v_n\{2\}$ for charged hadrons in Au+Au collisions at 200 GeV compared with experimental data from the STAR Collaboration \cite{Adamczyk:2016exq,Adamczyk:2017hdl}.}
  \label{fig3}
\end{figure}

The two-particle anisotropic flow coefficients $v_n\{2\}$ in Au+Au collisions are compared with the STAR measurement \cite{Adamczyk:2016exq,Adamczyk:2017hdl} in Fig.~\ref{fig3}. We multiply a factor of $N_{\rm part}$ with the $v_n\{2\}^2$ to scale out the natural dilution of correlation due to the increase in particle pairs while going from peripheral to central events. The IP-Glasma initial conditions with hydrodynamic evolution and coupling to UrQMD can reproduce the flow power spectrum from $v_2\{2\}$ to $v_5\{2\}$ from the most central to the $\sim40$\% centrality bin or for all $N_\mathrm{part} \gtrsim 100$.
%
%=======================================
%=======================================

%
%=======================================
\begin{figure}[ht!]
  \centering
   \includegraphics[width=0.95\linewidth]{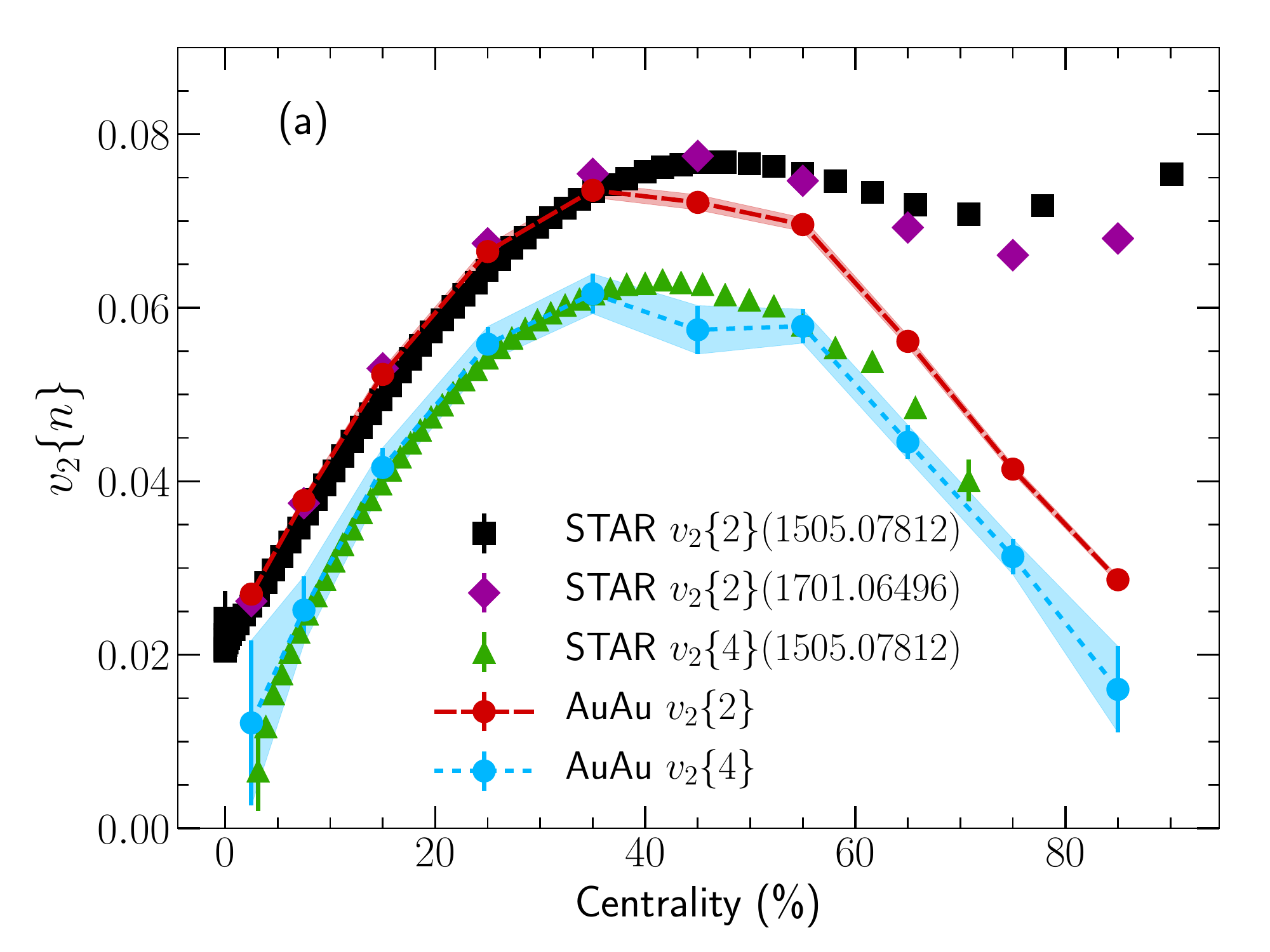}
   \includegraphics[width=0.95\linewidth]{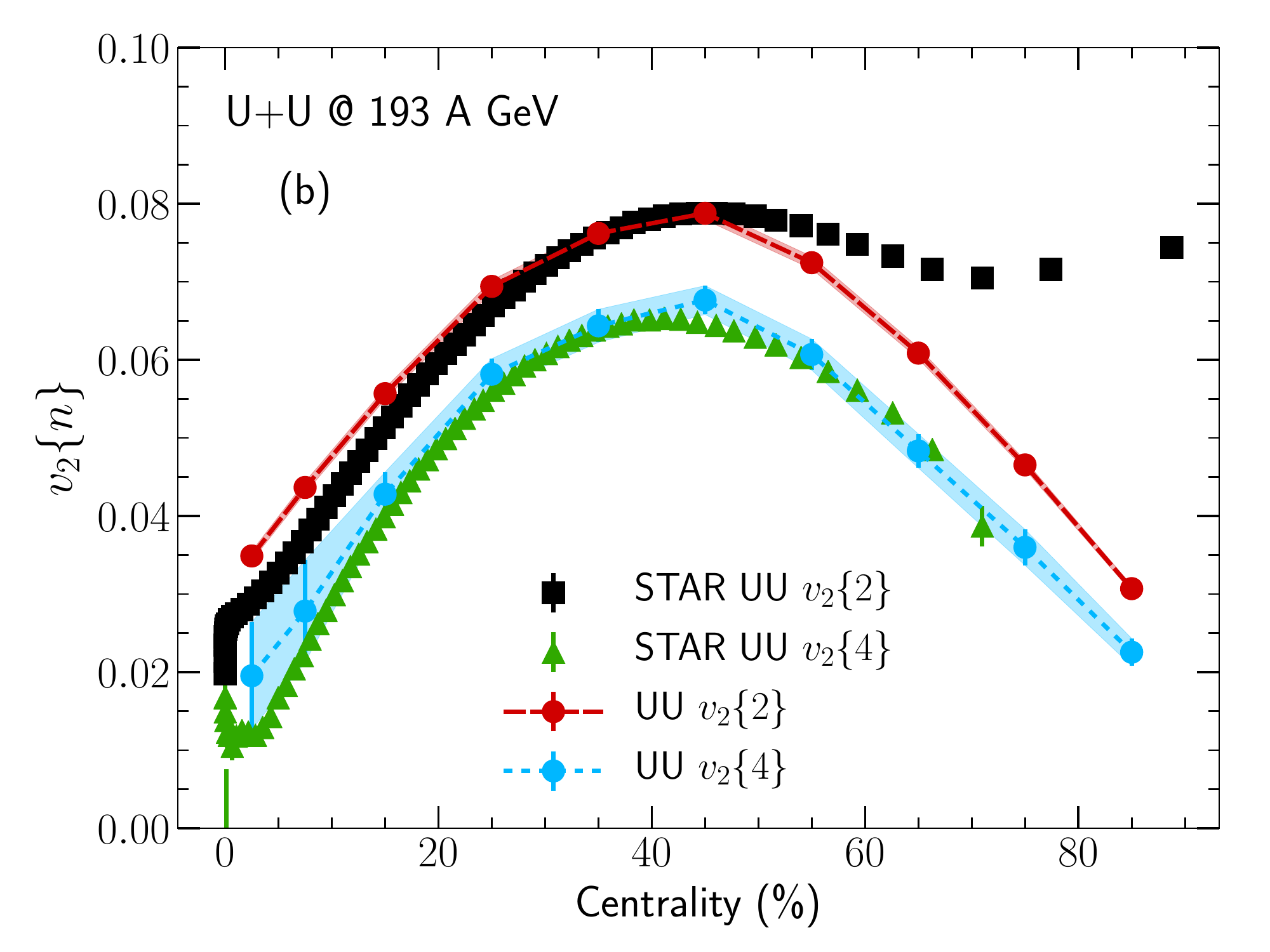}
  \caption{The two-particle and four-particle cumulant elliptic flow coefficient $v_2$ in Au+Au collisions (a) and U+U collisions (b) at RHIC, compared with experimental data from the STAR Collaboration \cite{Adamczyk:2017hdl,Adamczyk:2015obl}.}
  \label{fig4}
\end{figure}
%=======================================
%

For peripheral collisions beyond 40\% centrality, our calculations begin to underestimate the two-particle cumulant $v_n\{2\}$. The deviation from data can be seen more clearly in Fig.~\ref{fig4}a for the $v_2\{2\}$ results. In the same plot, we also compare our calculations with the 4-particle cumulant measurements. Unlike the comparison for the two-particle cumulant, good agreement of $v_2\{4\}$ with the STAR data was found across all centralities. This suggests that the deviation in the two-particle measurements is from non-flow correlations, such as correlations from di-jets that dominate in the peripheral events. This result nicely demonstrate the advantages of correlation measurements with more than two particles for extracting the transport properties of the QGP.

Fig.~\ref{fig4}b shows the comparison between theory predictions and the two $v_2$ measurements in U+U collisions for the same transport parameters as used for Au+Au collisions. The good agreement for $v_2\{4\}$ shows the quantitative predictive power of the hybrid theoretical framework. The slight over-estimation of the data for the most peripheral events, in particular for $v_2\{2\}$ could be because of the different bin sizes used in the calculation (5\% bins for the most central events) compared to the experiment, which uses much narrower bins. 

%
%=======================================
\begin{figure}[ht!]
  \centering
  \includegraphics[width=0.95\linewidth]{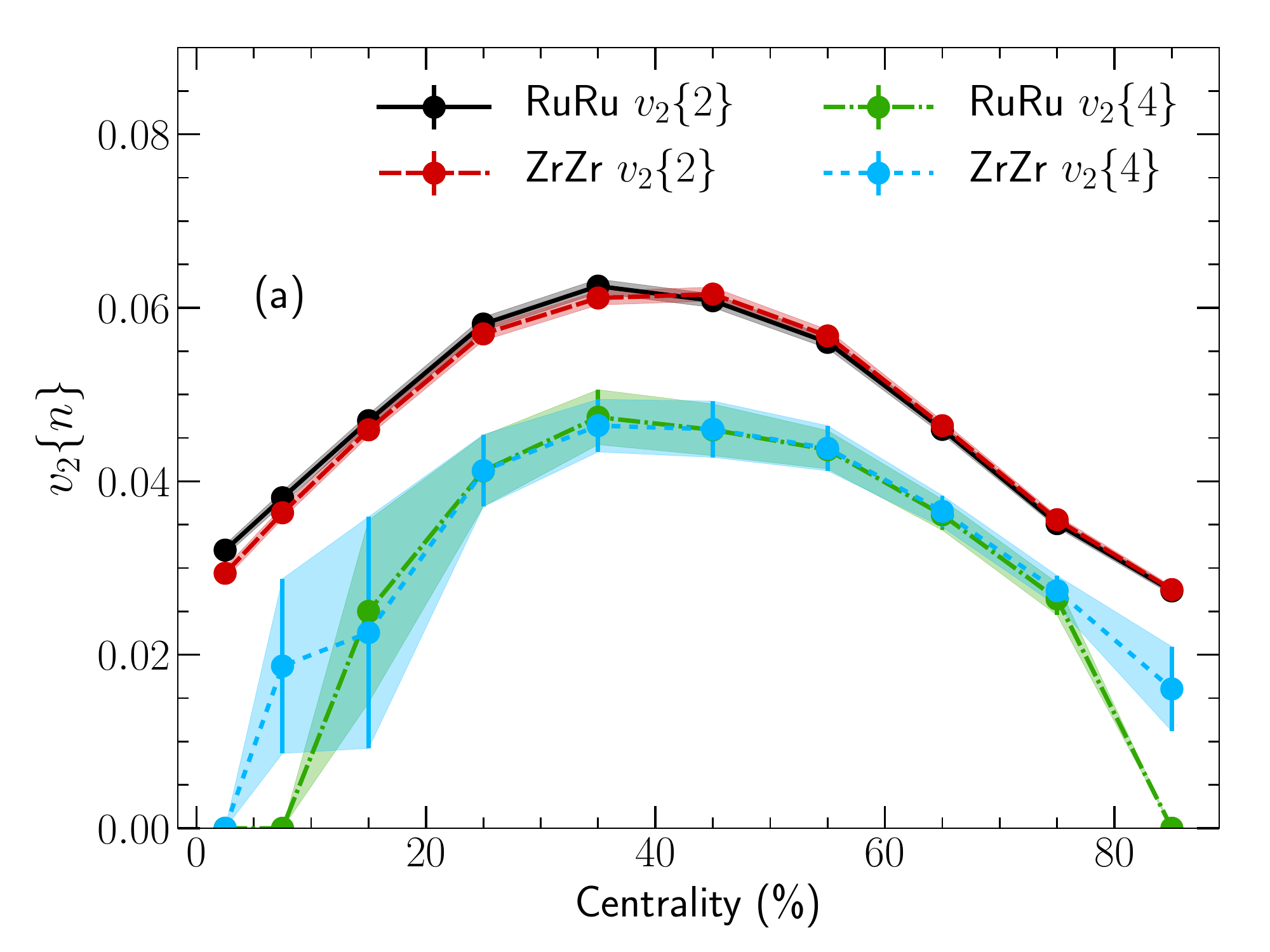} 
   \includegraphics[width=0.95\linewidth]{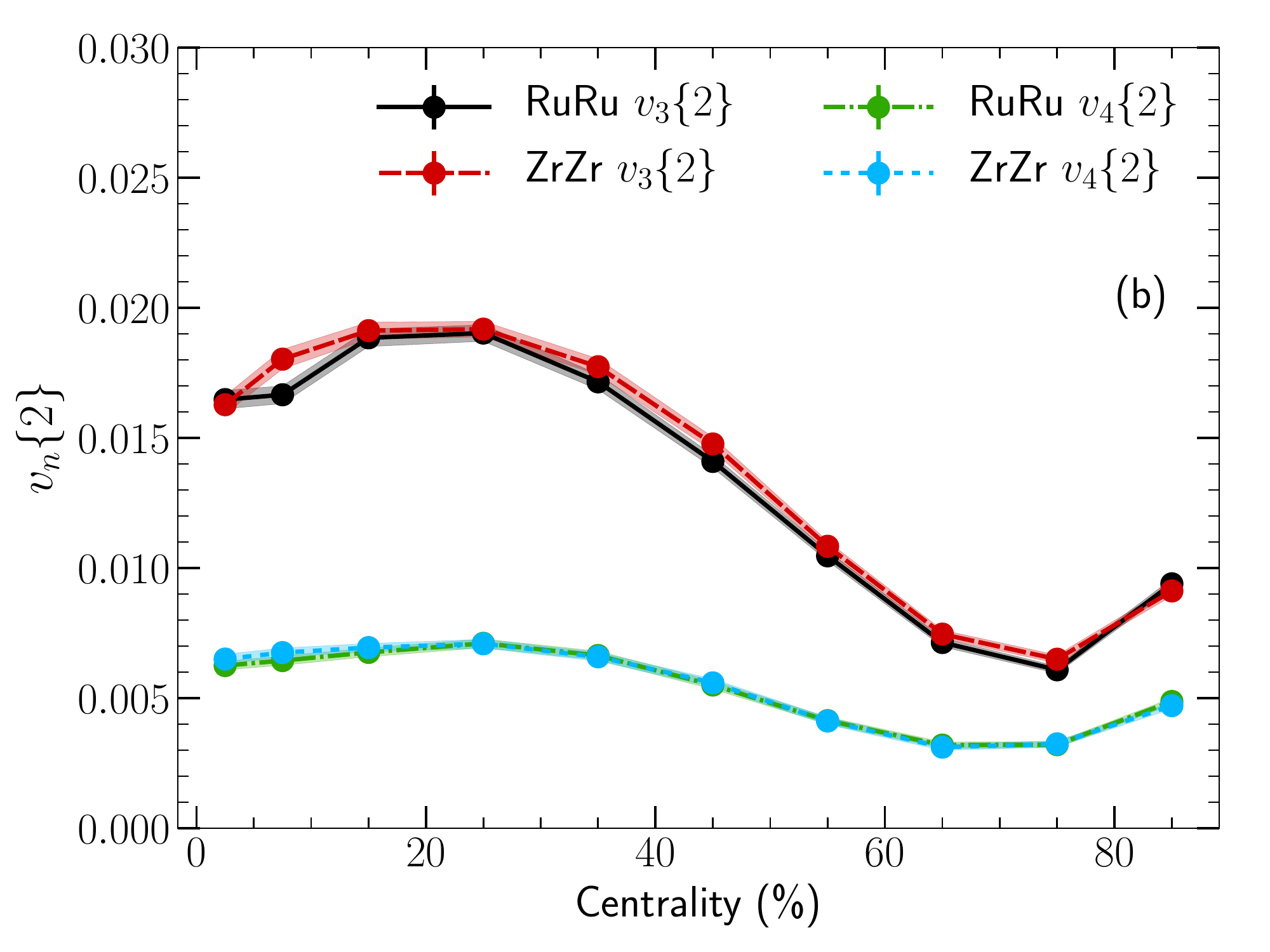}
  \caption{Panel (a): Prediction for the two-particle and four-particle cumulant $v_2\{2\}$ and $v_2\{4\}$ in Ru+Ru and Zr+Zr collisions at 200 GeV. Panel (b): Prediction for the higher order harmonic flow coefficients $v_{3,4}\{2\}$ in Ru+Ru and Zr+Zr collisions.}
  \label{fig5}
\end{figure}
%=======================================
%
Having achieved a good descriptions of the flow observables in Au+Au and U+U collisions, we move on to make predictions for the flow coefficients $v_n\{2\}$ and $v_n\{4\}$ in the significantly smaller Ru+Ru and Zr+Zr collisions in Fig.~\ref{fig5}. The comparison with the upcoming measurements in the RHIC isobar run will further test the predictive power of our framework, across different collision systems and geometries. 

%
%=======================================
\begin{figure*}[ht!]
  \includegraphics[width=0.65\linewidth]{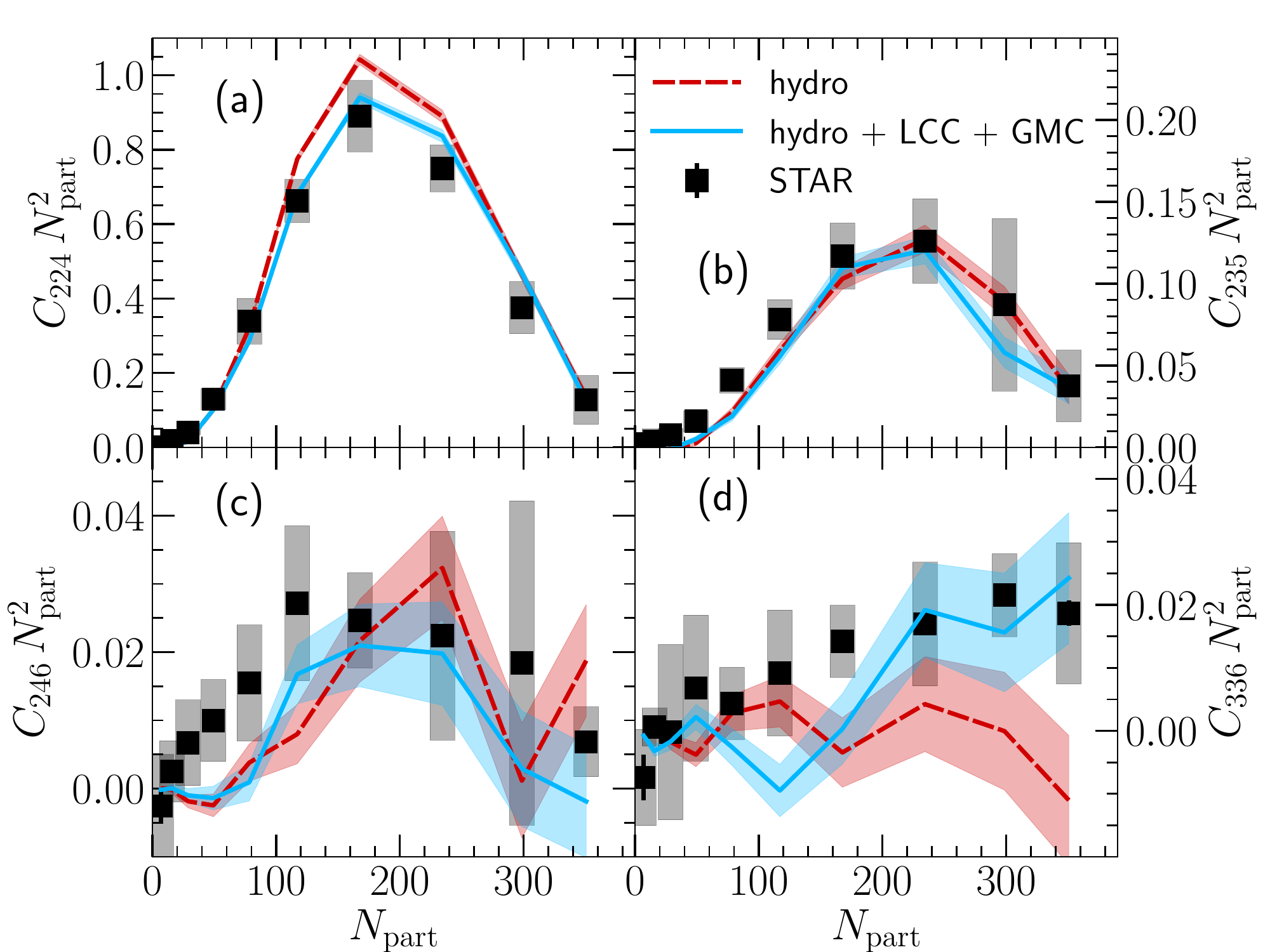}
  \caption{The three-particle correlators involving higher order harmonics compared with experimental data from the STAR Collaboration for Au+Au collisions at 200 GeV \cite{Adamczyk:2017hdl}.}
  \label{fig6}
\end{figure*}
%=======================================
%
In Fig.~\ref{fig5}a, we find approximately identical results for the two collision systems, except for a 5-10\% larger $v_2$ in the Ru+Ru system for the most central collisions. This is expected, because of the elliptical deformation introduced for the Ru nucleus. 

Predictions for higher harmonics in the two systems are presented in Fig.~\ref{fig5}b. Again, results are very similar between the two systems. There is a slight trend of larger $v_3$ in Zr+Zr compared to Ru+Ru for all centralities, except for the most central (0-5\%) and most peripheral (80-90\%) centrality class.

We attribute the rise of $v_3$ and $v_4$ in the most peripheral events to large effects from bulk viscosity, especially viscous corrections to the distribution function, in these very small systems with large gradients. We expect our results to only be robust up to centralities of 60-70\%. 

\subsection{Three-particle correlations}

In this section we study charge inclusive three particle correlations, which help us to understand the event-by-event correlation between different orders of flow harmonics. As we discuss later, introduction of charge dependence to these observables enables them to become sensitive to interesting phenomena such as the CME. 

We study the quantities
\begin{equation}\label{eq:Cmnk}
    C_{mnk} = \frac{\langle \Re\{Q_m Q_n Q_k^*\} \rangle_\mathrm{ev}}{\langle M(M-1)(M-2) \rangle_\mathrm{ev}}\,,
\end{equation}
where the $\langle \cdots \rangle_\mathrm{ev}$ denotes average over collision events and the $\Re\{\cdots\}$ stands for taking the real part of the enclosed complex variable. The $n$-th order particle flow vector $Q_n$ is defined as
\begin{equation}
    Q_n = \sum_{j=1}^{M} \tilde{w}_j e^{in\phi_j}\,.
\end{equation}
The sum is over all $M$ particles in the event (in a given kinematic range), and the weights $\tilde{w}_j$ are chosen as 1 in our calculation. The real and imaginary parts of $Q_n$ can be understood as the two components of a vector.

Eq.\,(\ref{eq:Cmnk}) can be expressed as
\begin{equation}
    C_{mnk} = \langle \cos(m\phi_1+n\phi_2-k\phi_3) \rangle \,,
\label{eq_cmnk}
\end{equation}
where the correlation function is averaged over the pairs and over collision events in one centrality, $\langle {\cal O} \rangle \equiv \frac{\sum_\mathrm{ev} \sum_\mathrm{pairs} {\cal O}}{\sum_\mathrm{ev} N_\mathrm{pairs}}$.
Only three particle correlations with $k=m+n$ are non-zero after event average \cite{Bhalerao:2011yg,Jia:2014jca}. Such correlators provide unique ways to study the initial state geometry and non-linear hydrodynamic response of the medium~\cite{Teaney:2010vd, Qiu:2012uy, Jia:2012ma,Teaney:2013dta,Bhalerao:2013ina,Niemi:2015qia,Giacalone:2016afq,Gardim:2016nrr,Aad:2014fla}.

If the azimuthal correlations are fully driven by hydrodynamics, Eq.\,\eqref{eq_cmnk} can be expressed in terms of flow harmonics and the corresponding event-planes as
\begin{equation}
    C_{mnk}=\langle v_m v_n v_k \cos(m\Psi_m+n\Psi_n-k\Psi_k) \rangle_{\rm hydro}.
    \label{eq_cmnk_hydro}
\end{equation}

However, such decomposition does not hold in general because of correlations that are of non-hydrodynamic origin, also explored in this paper. For example, if one of the harmonic $m,n=1$, then factorization breaking due to momentum conservation will lead to violation of Eq.\,\eqref{eq_cmnk_hydro}~\cite{Borghini:2002mv,Bzdak:2010fd}. Three particle correlators involving $m=n=1$ and their charge dependence were originally proposed to study two-particle correlations with respect to a specific event plane in the search for the CME signal~\cite{Voloshin:2004vk}. In the following section we study several correlators relevant for the CME search, providing a baseline for the case of no signal.  

Fig.~\ref{fig6} shows comparisons between the theory calculations and three particle correlations measured by the STAR Collaboration \cite{Adamczyk:2017hdl}. A factor of $N_{\rm part}^2$ is multiplied to scale out the trivial dilution of correlation with the increase of triplets. Good agreement is found for all centrality bins. This reflects that the correlation in initial eccentricities $\{\epsilon_n\}$ and the following hydrodynamic evolution can capture the correlation among flow harmonics of different orders. We point out that a significant fraction of the three-particle correlation is developed in the hadronic cascade phase. The pure hydro simulation results, compared to experimental data in Ref.~\cite{Adamczyk:2017byf}, underestimated many of the measured correlation functions. The effect of local charge conservation and global momentum conservation, visible as the difference between the solid (including both effects) and dashed (without either effect) lines, is weak for these observables. 

\begin{figure}[ht!]
  \includegraphics[width=0.95\linewidth]{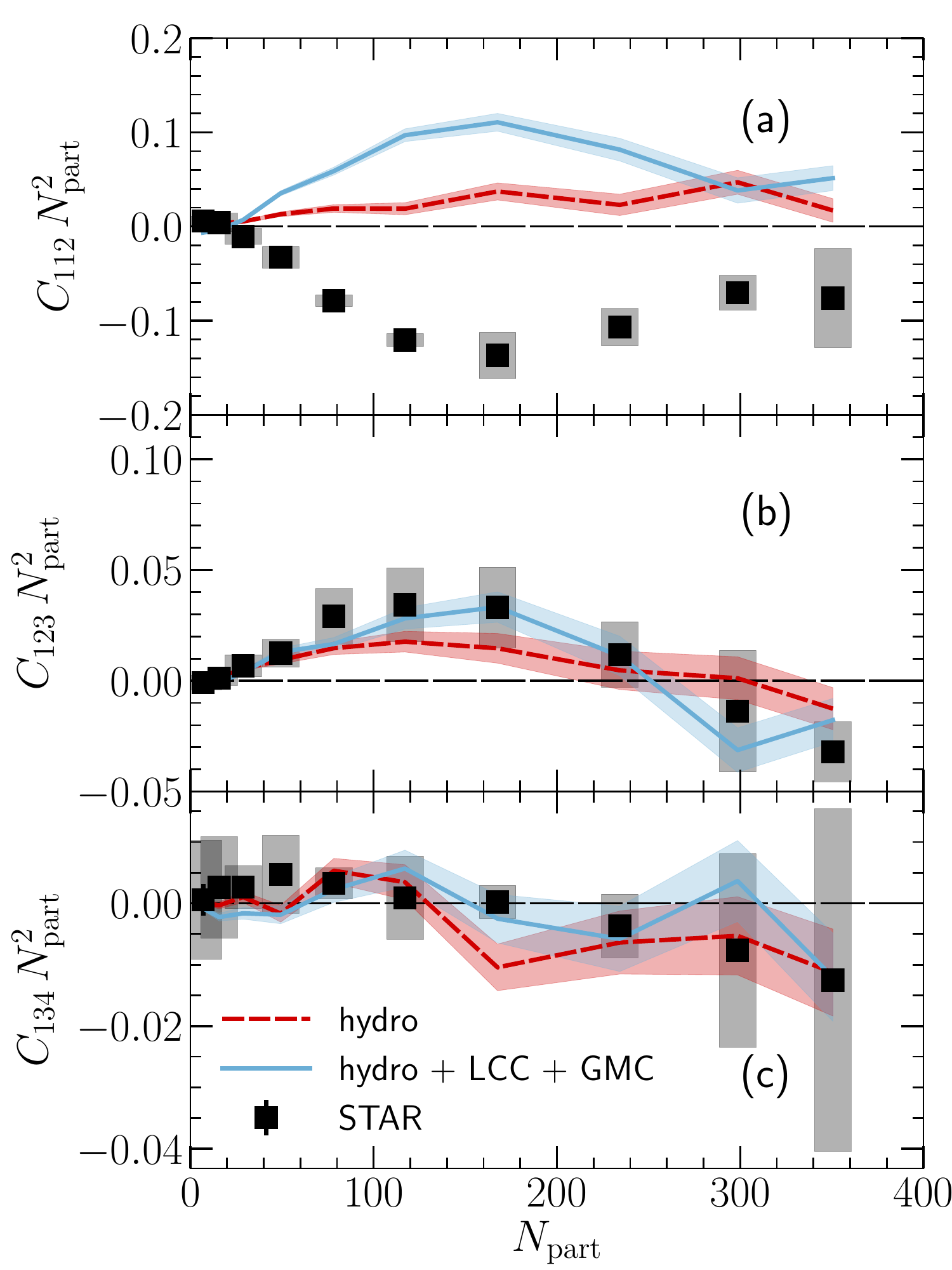}
  \caption{The three-particle correlators involving the first order harmonic compared with experimental data from the STAR Collaboration for Au+Au collisions at 200 GeV \cite{Adamczyk:2017hdl}.}
  \label{fig7}
\end{figure}

Correlators shown in Fig.\,\ref{fig6} do not include $v_1$ contributions, which are most sensitive to the effects of momentum conservation. Fig.\,\ref{fig7} shows a compilation of $C_{mnk}$ for $m=1$ and various other combinations of $k=n+1$. For $n=2$ and $n=3$ agreement with experimental data from the STAR Collaboration \cite{Adamczyk:2017hdl} is good, however, for $n=1$, we see a large difference between the calculation and the experimental data. Even the sign of $C_{112}$ is calculated to be opposite to the experimentally determined one. The effect of global momentum conservation brings the result closer to the experimental data, but the sign is still opposite to the experimental result. It is likely that a mechanism of local momentum conservation during particle sampling will improve the agreement with the experimental data. Because both $m$ and $n$ are equal one here, sensitivity to exact momentum conservation is potentially very strong for $C_{112}$. We will investigate a possible extensions of our framework in this direction in the future. 
%
%=======================================
%=======================================
%

\subsection{Four-particle Symmetric Cumulants}

We conclude the discussion of charge inclusive multi-particle correlations by showing the results for four-particle symmetric cumulants \cite{Bilandzic:2013kga}, which reveal the amount of correlation between the magnitudes of different orders of flow harmonics.

The symmetric cumulant measurements are defined as,
\begin{eqnarray}
    SC\{m,n\} &=& \frac{\langle Q_n Q_n^* Q_m Q_m^* \rangle_\mathrm{ev}}{\langle M(M-1)(M-2)(M-3) \rangle_\mathrm{ev}} \notag \\
    && - \frac{\langle Q_n Q_n^*\rangle_\mathrm{ev}}{\langle M(M-1) \rangle_\mathrm{ev}} \frac{\langle Q_m Q_m^* \rangle_\mathrm{ev}}{\langle M(M-1) \rangle_\mathrm{ev}}\,.    
\end{eqnarray}
Self-correlations in the 2-particle and 4-particle correlation functions are subtracted. To eliminate the effect of the magnitudes of $v_m$ and $v_n$ on the value of the symmetric cumulant, we divide by their average values and define the normalized symmetric cumulant
\begin{equation}
    NSC\{m,n\} = \frac{SC\{m,n\}}{\frac{\langle Q_n Q_n^*\rangle_\mathrm{ev}}{\langle M(M-1) \rangle_\mathrm{ev}} \frac{\langle Q_m Q_m^*\rangle_\mathrm{ev}}{\langle M(M-1) \rangle_\mathrm{ev}}}.
\end{equation}

\begin{figure}[h!]
  \includegraphics[width=0.95\linewidth]{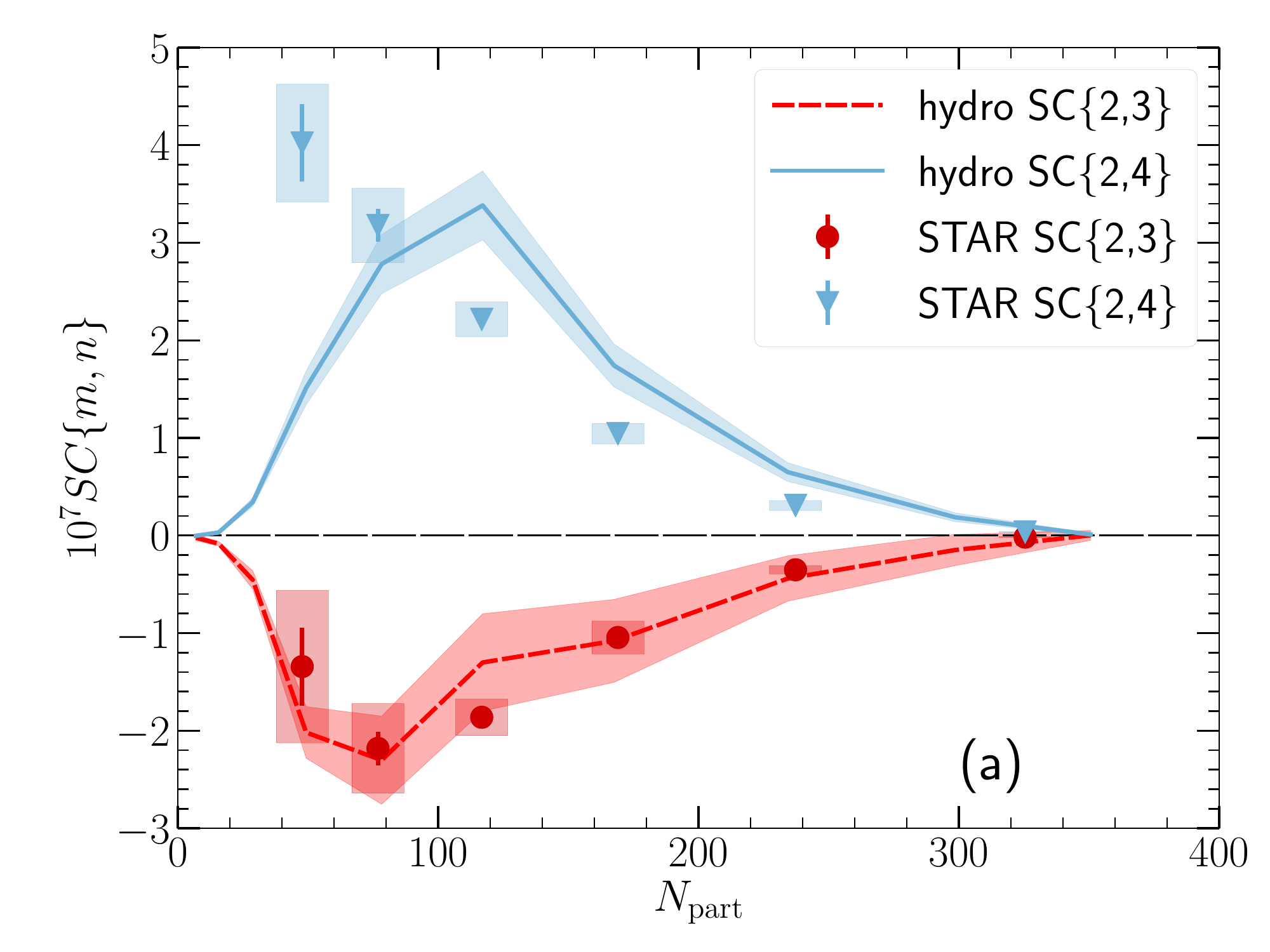}
  \includegraphics[width=0.95\linewidth]{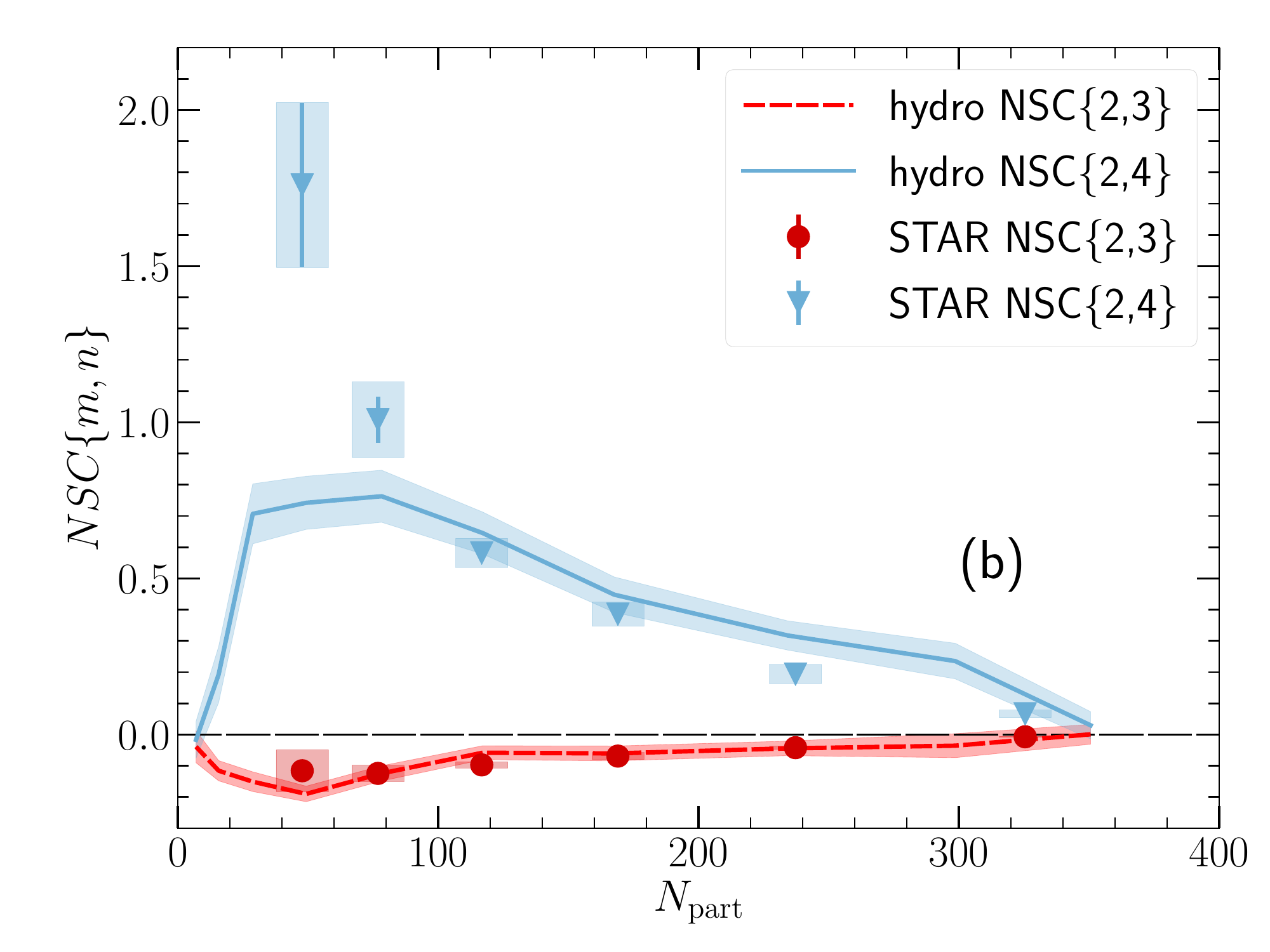}
  \caption{The 4-particle symmetric cumulants (a) and normalized symmetric cumulants (b) compared with experimental data from the STAR Collaboration in Au+Au collisions at 200 GeV \cite{STAR:2018fpo}.}
  \label{fig8}
\end{figure}

We present results for $SC\{2,3\}$ and $SC\{2,4\}$ in Fig.\,\ref{fig8}a and compare to data from the STAR Collaboration \cite{STAR:2018fpo}. We reproduce the experimentally observed anti-correlation between $v_2$ and $v_3$, encoded in the negative sign of $SC\{2,3\}$, as well as its centrality dependence. Agreement of our result for $SC\{2,4\}$ with experimental data is reasonable for $N_{\rm part}>100$, with the calculation being systematically above the data, while at small $N_{\rm part}$ we underestimate the experimental result. 

The normalized symmetric cumulants presented in Fig.\,\ref{fig8}b show the same behavior but emphasize that the anti-correlation between $v_2$ and $v_3$ is rather weak compared to the correlation between $v_2$ and $v_4$.

%=======================================
%=======================================
%

\begin{table}[h!]
\begin{center}
    \begin{tabular}{ | c | c | c |}
    \hline
          & $10^{7}SC\{2,3\}$ & $10^{7}SC\{2,4\}$  \\ \hline
    decay only & $-0.3(4)$  & $0.9(1)$  \\ \hline
    with UrQMD & $-1.1(4)$ & $1.7(2)$  \\
    \hline
    \end{tabular}
\end{center}
\caption{The effects of late stage hadronic transport on the symmetric cumulant in 20-30\% Au+Au collisions at 200 GeV. The numbers in the parenthesis show the statistical errors on the last digit. \label{tab:UrQMD_on_SC}}
\end{table}

We note that also for the symmetric cumulants the main reason for better agreement with the experimental data compared to hydrodynamic calculations \cite{Gardim:2016nrr} presented in \cite{STAR:2018fpo} is the inclusion of the hadronic afterburner (see Table~\ref{tab:UrQMD_on_SC}). The larger effective viscosities we use also tend to increase the correlations between different harmonics, but likely have only a weak effect \cite{Gardim:2016nrr}.
The rapid increase of the measured $SC\{2,4\}$ towards small $N_\mathrm{part}$ suggests that at low multiplicity residual non-flow effects present even in four-particle correlations begin to dominate the experimental result~\cite{Huo:2017nms}.

\section{Charge dependent observables} \label{sec:charge}

The charge dependent two-particle and three-particle correlation functions are defined as
\begin{eqnarray}
    C_n\{2\} &=& \langle \cos(n (\phi_\alpha - \phi_\beta))\rangle\,,\\
    C_{mnk} &=& \langle \cos(m \phi_\alpha + n \phi_\beta - k \phi_c)\rangle\,,
\end{eqnarray}
where $\phi_\alpha$ and $\phi_\beta$ can be assigned different charges in the following way:
\begin{eqnarray}
C_{n}\{2\}(\mathrm{OS}) &=&\langle \cos(n (\phi_\alpha^\pm - \phi_\beta^\mp))\rangle\,,\\
C_{n}\{2\}(\mathrm{SS}) &=&\langle \cos(n (\phi_\alpha^\pm - \phi_\beta^\pm))\rangle\,,\\
C_{mnk}(\mathrm{OS}) &=& \langle \cos(m \phi_\alpha^{\pm} + n \phi_\beta^{\mp} - k \phi_c)\rangle\,,\\
C_{mnk}(\mathrm{SS}) &=& \langle \cos(m \phi_\alpha^{\pm} + n \phi_\beta^{\pm} - k \phi_c)\rangle\,.
\end{eqnarray}
Here $\langle \cdots \rangle$ denotes average over all pairs or triplets within a given centrality bin. In $C_{mnk}$, the third particle $\phi_c$ is chosen to be charge inclusive. The notations ``OS" and ``SS" refer to opposite-sign and same-sign correlators respectively. 

In the following, we consider the specific case of events with zero net charge. The numbers of positive and negative particles in one collision event are $N_+ = N_- = N_\mathrm{ch}/2$, where $N_\mathrm{ch}$ is the number of charged hadrons.

For two particle correlation functions, the number of opposite-sign and same-sign pairs are $N_\mathrm{OS} = 2 N_+ N_- = N_\mathrm{ch}^2/2$ and $N_\mathrm{SS} = N_+(N_+-1)+N_-(N_--1) = N^2_\mathrm{ch}/2 - N_\mathrm{ch}$ respectively. The difference between opposite-sign and same-sign two-particle correlation functions can be written as,
\begin{eqnarray}
    \Delta C_n\{2\} &=& C_n\{2\}(\mathrm{OS}) - C_n\{2\}(\mathrm{SS}) \notag \\
     &&   \hspace{-1cm} = \frac{\sum_\mathrm{OS}\cos(n(\phi_\alpha - \phi_{\beta}))}{N_\mathrm{OS}} - \frac{\sum_\mathrm{SS}\cos(n(\phi_\alpha - \phi_{\beta}))}{N_\mathrm{SS}}\, \notag \\
        && \hspace{-1cm} = \frac{1}{N_\mathrm{OS}} \bigg[\sum_\mathrm{OS} \cos(n(\phi_\alpha - \phi_{\beta})) - \sum_\mathrm{SS}\cos(n(\phi_\alpha - \phi_{\beta})) \notag \\
        && \quad - \frac{N_\mathrm{OS} - N_\mathrm{SS}}{N_\mathrm{SS}} \sum_\mathrm{SS} \cos(n(\phi_\alpha - \phi_\beta)) \bigg]\,.
\label{eq:DeltaC1_2}
\end{eqnarray}

\begin{figure}[ht!]
  \includegraphics[width=1.0\linewidth]{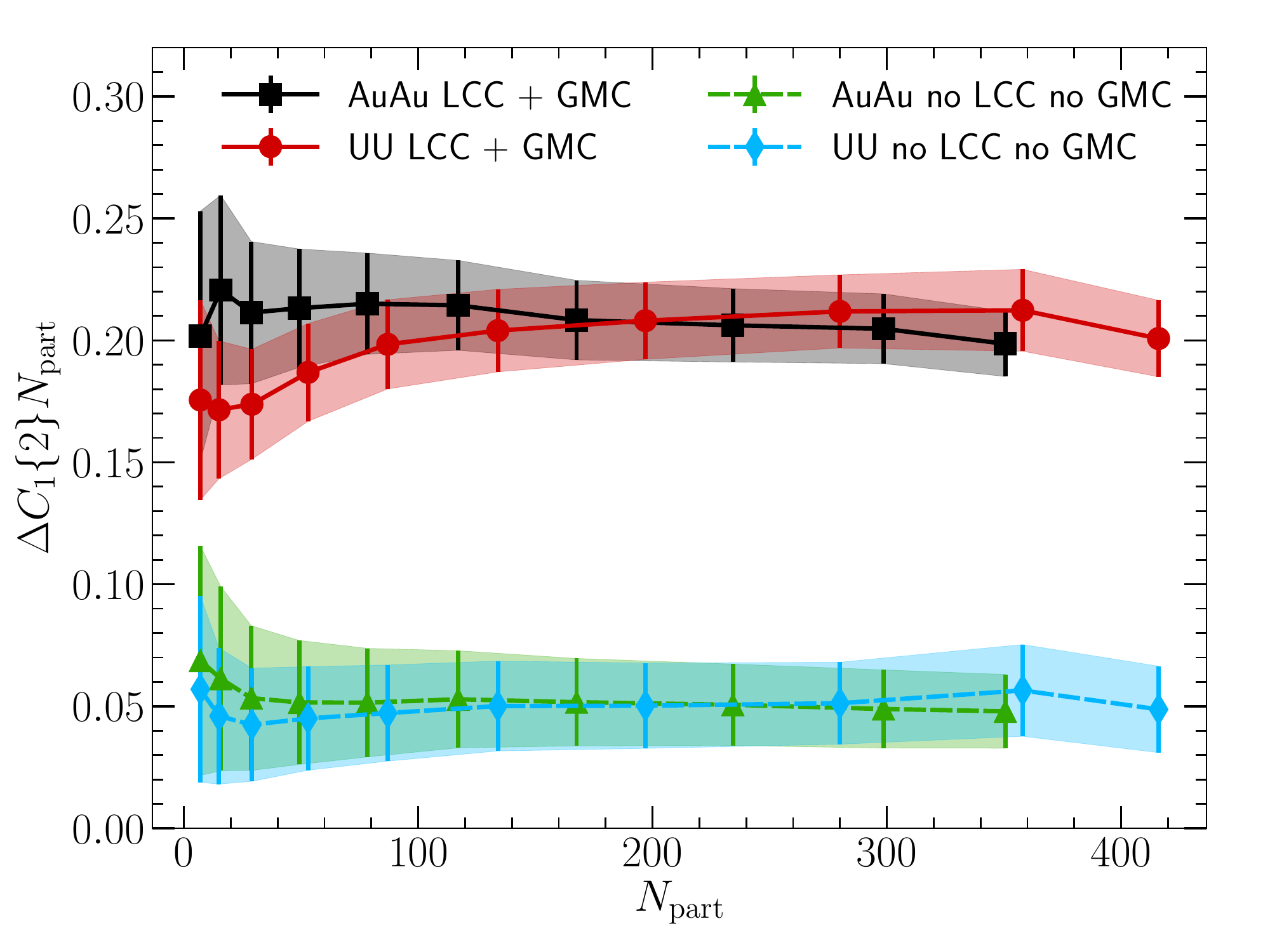}
  \caption{The difference between opposite-sign and same-sign two-particle correlation functions $\Delta C_1\{2\} = C_1\{2\}(\mathrm{OS}) - C_1\{2\}(\mathrm{SS})$ scaled by number of participants $N_\mathrm{part}$ in Au+Au and U+U collisions. Results with and without imposing local charge conservation (LCC) and global momentum conservation (GMC) are shown.}
  \label{fig9}
\end{figure}

\begin{figure}[ht!]
  \includegraphics[width=1.0\linewidth]{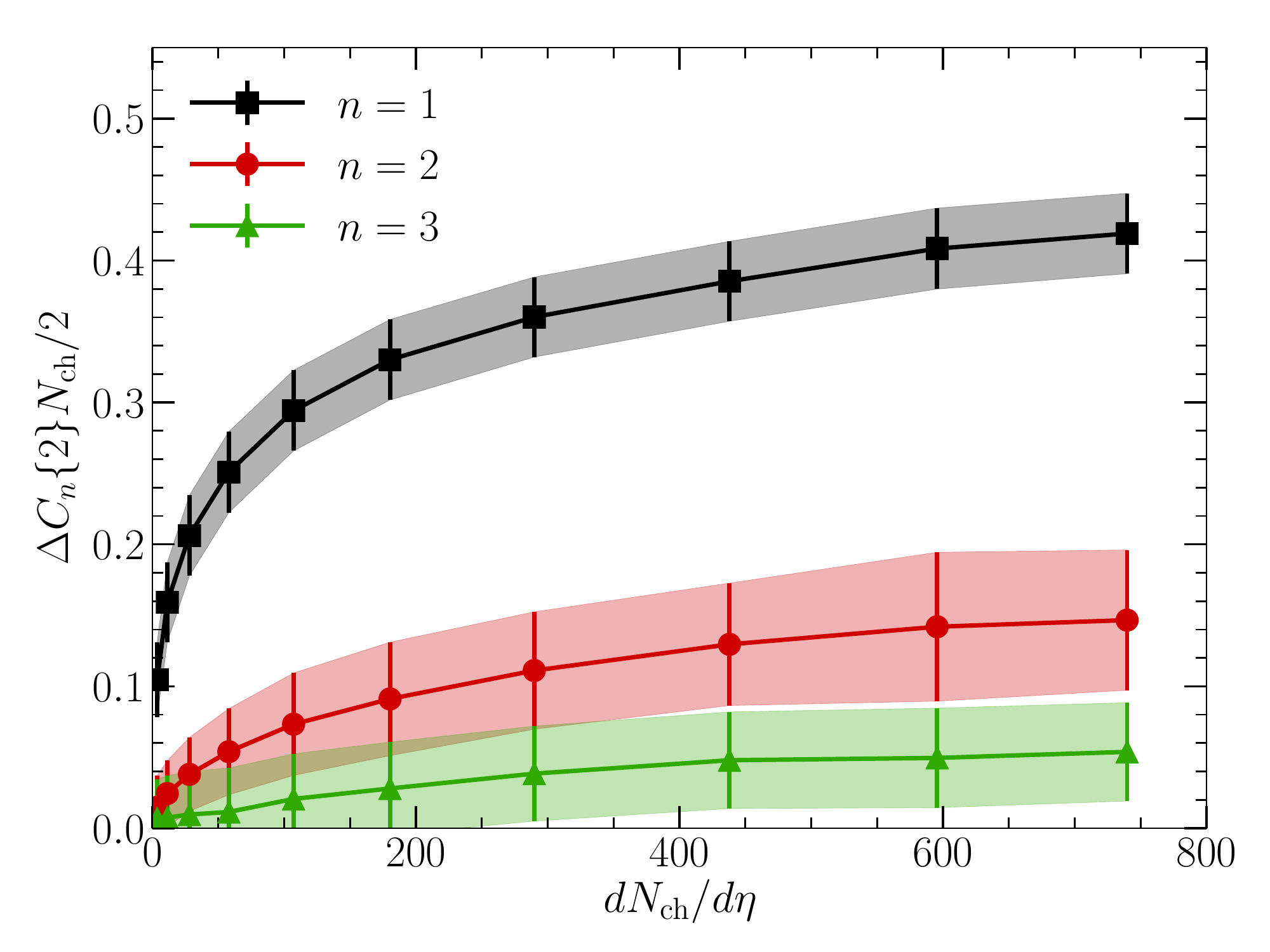}
  \caption{The difference between opposite-sign and same-sign two-particle correlation functions $\Delta C_n\{2\} = C_n\{2\}(\mathrm{OS}) - C_n\{2\}(\mathrm{SS})$ scaled by the number of charged particles $N_\mathrm{ch}$ in the rapidity interval $-1<\eta<1$ divided by 2 in Au+Au collisions at 200 GeV.}
  \label{fig9Nch}
\end{figure}

In our model, the positive and negative particles are produced from the same flow background. Thus, the flow background signals and the effect of global momentum conservation are expected to be cancelled in the difference between the first two terms of Eq.\eqref{eq:DeltaC1_2}, which then only contains contributions from truly correlated pairs:
\begin{eqnarray}
    &&\sum_\mathrm{OS} \cos(n(\phi_\alpha - \phi_{\beta})) - \sum_\mathrm{SS}\cos(n(\phi_\alpha - \phi_{\beta})) \notag\\ 
    && \quad=\!\! \sum_\mathrm{corr.\,pairs} \!\!\!\!\cos(n(\phi_\alpha - \phi_{\alpha'})),
\end{eqnarray}
where the summation on the R.H.S. runs over all the correlated positive and negative charge pairs with angles $\phi_\alpha$ and $\phi_{\alpha'}$, respectively. 
Two sources of such correlated charged pair production in our framework are: 1) decay of neutral resonances and 2) local charge conservation in the sampling process.
The last term in Eq.~(\ref{eq:DeltaC1_2}) exists because of unequal numbers of opposite and same sign pairs, and can be written as 
\begin{equation}
    \frac{N_\mathrm{OS} - N_\mathrm{SS}}{N_\mathrm{SS}} \sum_\mathrm{SS} \cos(n(\phi_\alpha - \phi_\beta)) = N_\mathrm{ch} C_n\{2\}(\mathrm{SS}).
\end{equation}
Thus Eq.~(\ref{eq:DeltaC1_2}) can be further simplified as,
\begin{equation}
    \Delta C_n\{2\} = \frac{2}{N_\mathrm{ch}} \bigg[K_n - C_n\{2\}(\mathrm{SS}) \bigg]\,,
\label{eq:Delta_C1_2_f}
\end{equation}
where
\begin{equation}
   K_n \equiv \langle \cos(n(\phi_\alpha - \phi_{\alpha'})) \rangle =  \frac{1}{N_{\rm ch}} \sum_\mathrm{corr.\,pairs} \!\!\!\!\cos(n(\phi_\alpha - \phi_{\alpha'}))\,.
\label{eq:K_n}
\end{equation}
Eq.~(\ref{eq:Delta_C1_2_f}) suggests that the $\Delta C_n\{2\}$ are inversely proportional to the charged hadron multiplicity. This is observed in our numerical results as shown in Fig.\,\ref{fig9} for $\Delta C_1\{2\}$, where we use $N_{\rm part}$ as a proxy for $N_{\rm ch}$. For central Au+Au events at top RHIC energy $dN_{\rm ch}/d\eta \approx 2 N_{\rm part}$. Because we evaluate $\Delta C_1\{2\}$ within two units of pesudorapidity (-1<$\eta$<1) one should in principle scale it by $N_{\rm ch}=2 dN_{\rm ch}/d\eta \approx 4 N_{part}$ to get the correct magnitude of $\langle \cos(\phi_\alpha - \phi_{\alpha'}) \rangle$.
Fig.~\ref{fig9Nch} further shows higher order $\Delta C_n\{2\} N_\mathrm{ch}/2$ for $n = 1-3$ in Au+Au collisions. The magnitude of $\Delta C_n\{2\}$ decreases rapidly with increasing order $n$.

%
%=======================================
%=======================================
%
%
%=======================================
%=======================================
%

With the implementation of local charge conservation, the net charge in every event is zero. All the positive particles have their own negative partners. The particle $\alpha$ and its pair partner $\alpha'$ have a short-range correlation in their rapidities and azimuthal angles $\phi_\alpha \sim \phi_{\alpha'}$, caused by the focusing due to the collective flow. We set $\phi_{\alpha'} = \phi_\alpha - \delta \phi_\alpha$, where the collection of the angles $\{\delta \phi_\alpha\}$ can fluctuate symmetrically around 0 for different pairs. For a narrow distribution of $\{\delta \phi_\alpha \}$, we would expect $K_1 \sim 1$, which is much larger than the second term $C_1\{2\}(\mathrm{SS})$ in Eq.~(\ref{eq:Delta_C1_2_f}). 

As shown in Fig.\,\ref{fig9}, the difference between the cases including LCC and those that do not, is large, approximately a factor of four. Based on above discussion, this can be attributed to the increase of $K_1 = \langle \cos(\delta\phi_\alpha) \rangle$ when particle pairs are forced to be produced in the same cell and experience the same local boost from flow, as per our implementation of LCC. The finite value of $\Delta C_1\{2\}$ in the case of no LCC implementation is a result of charge dependence in the various resonance decays and interactions in the hadronic afterburner. Another important observation is that within the uncertainties we do not see any significant system dependence for Au+Au and U+U.

The fact that the $\Delta C_n\{2\}N_{\rm ch}/2$ are significantly smaller than 1 indicates that the focusing by collective flow is too weak to generate a very strongly peaked distribution of $\delta\phi_\alpha$ around zero. Experimental measurements of different orders of $\Delta C_n$ at RHIC will well constrain the width of the $\delta\phi_\alpha$ distribution. We note that, interestingly, the result for $\Delta C_1\{2\}$ scales well with $N_{\rm part}$ and less so with $N_{\rm ch}$, even though our approximate expression \eqref{eq:Delta_C1_2_f} predicts scaling with $N_{\rm ch}$.

We now investigate if a similar centrality or system dependence is expected for $\Delta C_{mnk}$ in our model. This is important since the signal of the CME is expected to be driven by the magnetic field, which would lead to a strong system and centrality dependence of $\Delta C_{112}$ \cite{Voloshin:2010ut, Bzdak:2011yy, Bloczynski:2012en, Bloczynski:2013mca, Chatterjee:2014sea, Huang:2015oca}.

Following similar arguments as for the two-particle correlators, the charge dependence of the three-particle correlation functions can be written as (note that $k=m+n$)
\begin{eqnarray}
    \Delta C_{mnk} &=& C_{mnk}(\mathrm{OS})-C_{mnk}(\mathrm{SS}) \label{eq:Delta_Cmnk} \\
        &=&\frac{2}{N_\mathrm{ch}}\bigg[ \langle \cos(m\phi_\alpha + n \phi_{\alpha'} - k \phi_c) \rangle - C_{mnk}(\mathrm{SS}) \bigg]. \notag
\end{eqnarray}
Similar to the two-particle correlations, the second term $C_{mnk}(\mathrm{SS})$ is usually much smaller than the first term in Eq.~(\ref{eq:Delta_Cmnk}).
To simplify the first term, we can define new variables, $\phi_\mathrm{pair,\alpha}=(\phi_\alpha + \phi_{\alpha'})/2$ and $\delta \phi_\alpha = \phi_\alpha - \phi_{\alpha'}$. Then
\begin{eqnarray}
    && \!\!\! \langle \cos(m\phi_\alpha + n \phi_{\alpha'} - k \phi_c) \rangle \notag \\
    && \qquad = \left\langle \cos \left[k(\phi_\mathrm{pair,\alpha} - \phi_c) + \frac{m-n}{2} \delta \phi_\alpha \right] \right\rangle \notag \\
    && \qquad = \left\langle \cos[k(\phi_\mathrm{pair,\alpha} - \phi_c)] \cos \left( \frac{m-n}{2} \delta \phi_\alpha \right) \right\rangle \notag \\
    && \qquad\quad - \left\langle \sin[k(\phi_\mathrm{pair,\alpha} - \phi_c)] \sin \left( \frac{m-n}{2} \delta \phi_\alpha \right) \right\rangle
\label{eq:cos_mnk}
\end{eqnarray}
Using the reflection symmetry in particle pairs and the fact that $\phi_\mathrm{pair,\alpha}$ and $\delta \phi_\alpha$ are independent variables,
\begin{equation}
    \left\langle \sin[k(\phi_\mathrm{pair,\alpha} - \phi_c)] \sin \left( \frac{m-n}{2} \delta \phi_\alpha \right) \right\rangle = 0.
\end{equation}
We emphasize that this term only averages to zero when one decomposes the pair angles into $\phi_{\mathrm{pair}, \alpha}$ and $\delta \phi_\alpha$ in Eq. (\ref{eq:cos_mnk}). The decomposition of three particle correlators into two-particle correlations in the appendix of Ref. \cite{Sirunyan:2017quh}, where a different sine term is omitted, may not be valid in general.
Thus, Eq.~(\ref{eq:cos_mnk}) can be simplified to read
\begin{eqnarray}
    && \!\!\! \langle \cos(m\phi_\alpha + n \phi_{\alpha'} - k \phi_c) \rangle \notag \\
    && \qquad = \left\langle \cos[k(\phi_\mathrm{pair,\alpha} - \phi_c)] \cos \left( \frac{m-n}{2} \delta \phi_\alpha \right) \right\rangle.
\label{eq:cos_mnk_final1}
\end{eqnarray}
For the special case $m = n$,
\begin{eqnarray}
   \langle \cos(m\phi_\alpha + n \phi_{\alpha'} - k \phi_c) \rangle &=& \left\langle \cos[k(\phi_\mathrm{pair,\alpha} - \phi_c)] \right\rangle \notag \\
   &\equiv& C_k^\mathrm{pair}\{2\}.
\label{eq:cos_mnk_final2}
\end{eqnarray}
Here $C_k^\mathrm{pair}\{2\}$ denotes the $k$-th order harmonic coefficient of the two-particle correlation function between the correlated pair angle and charged hadrons.

Now for the context of CME related measurements, we concentrate on the charge dependence of  $C_{mnk}$ for first three lowest order harmonic combinations $C_{112}, C_{123}$ and $C_{132}$. By neglecting the second term in Eq.~(\ref{eq:Delta_Cmnk}) and using Eqs.~(\ref{eq:cos_mnk_final1}) and (\ref{eq:cos_mnk_final2}), we can derive
\begin{eqnarray}
    \Delta C_{112} \hspace{-0.05cm}&\approx&\hspace{-0.1cm} \frac{2}{N_\mathrm{ch}} \left\langle \cos[2(\phi_\mathrm{pair,\alpha} - \phi_c)] \right\rangle \hspace{-0.1cm}=\hspace{-0.1cm} \frac{2}{N_{\rm ch}} C_2^\mathrm{pair}\{2\} \label{eq:DeltaC112} \\
    \Delta C_{132} &\approx& \frac{2}{N_\mathrm{ch}} \left\langle \cos[2(\phi_\mathrm{pair,\alpha} - \phi_c)] \cos \left(2\delta \phi_\alpha \right) \right\rangle, \label{eq:DeltaC132} \\    
    \Delta C_{123} &\approx& \frac{2}{N_\mathrm{ch}} \left\langle \cos[3(\phi_\mathrm{pair,\alpha} - \phi_c)] \cos \left( \frac{\delta \phi_\alpha}{2} \right) \right\rangle. \label{eq:DeltaC123} 
\end{eqnarray}

Here, note the difference between the charge dependent three-particle correlator $C_{132} \equiv \langle \cos(\phi^{\mp}_\alpha - 3 \phi^{\pm}_\beta + 2 \phi_c) \rangle$ and $C_{123} \equiv \langle \cos(\phi^{\mp}_\alpha + 2 \phi^{\pm}_\beta - 3 \phi_c) \rangle$. 
If there is no correlation between $\phi_\mathrm{pair,\alpha}$ and $\delta \phi_\alpha$, the event-average in Eqs.~(\ref{eq:DeltaC132}) and (\ref{eq:DeltaC123}) can be factorized as,
\begin{eqnarray}
    \Delta C_{132} &=& \frac{2}{N_\mathrm{ch}} C_2^\mathrm{pair}\{2\} K_2, \label{eq:C132_factorized} \\    
    \Delta C_{123} &=& \frac{2}{N_\mathrm{ch}} C_3^\mathrm{pair}\{2\} K_{1/2}, \label{eq:C123_factorized} 
\end{eqnarray}
where $K_n$ is defined in Eq.~(\ref{eq:K_n}). However, this factorization is expected to be badly broken because of the hydrodynamic anisotropic flow. The elliptic flow generates a stronger boost for the pairs emitted along the second order event plane $\Psi_2$ compared to pairs emitted perpendicular to $\Psi_2$. This effect leads to a positive correlation between $\cos[2(\phi_\mathrm{pair,\alpha} - \phi_c)]$ and $\cos(2\delta \phi_\alpha)$ in Eq.~(\ref{eq:DeltaC132}). A similar positive correlation between $\cos[3(\phi_\mathrm{pair,\alpha} - \phi_c)]$ and $\cos(\delta \phi_\alpha/2)$ is generated from the underlying triangular flow in Eq.~(\ref{eq:DeltaC123}). 

Regardless of how well the factorizations in  Eq.~(\ref{eq:C132_factorized}) and Eq.~(\ref{eq:C123_factorized}) hold, the coefficient $C_k^\mathrm{pair}\{2\}$ is related to the charged hadron two-particle anisotropic coefficient $C_k\{2\}$. One can show that
\begin{eqnarray}
    C_k\{2\} &=& \frac{1}{2} \left\{\left\langle \cos[k(\phi_\alpha - \phi_c)] \right\rangle + \left\langle \cos[k(\phi_{\alpha'} - \phi_c)] \right\rangle\right\} \notag \\
    &=& \left\langle \cos[k(\phi_\mathrm{pair,\alpha} - \phi_c)] \cos \left( \frac{k}{2} \delta \phi_\alpha \right) \right\rangle. \label{Eq:Ck2pair}
\end{eqnarray}
Therefore, based on Eqs.\,\eqref{eq:DeltaC112}, \eqref{eq:DeltaC132}, and \eqref{Eq:Ck2pair}, we expect the following hierarchy,
\begin{equation}
    \frac{\Delta C_{112}}{C_2\{2\}} \ge \frac{\Delta C_{132}}{C_2\{2\}}.
\label{eq:Delta112_over_C2}
\end{equation} 
The equal sign is fulfilled only if $\delta \phi_\alpha = 0$ for all the correlated pairs. For the correlation function $\Delta C_{123}$ we have
\begin{equation}
    \frac{N_\mathrm{ch}}{2}\frac{\Delta C_{123}}{C_3\{2\}} = \frac{\left\langle \cos(3(\phi_\mathrm{pair,\alpha} - \phi_c)) \cos \left(\delta \phi_\alpha/2 \right) \right\rangle}{\left\langle \cos(3(\phi_\mathrm{pair,\alpha} - \phi_c)) \cos \left(\delta 3\phi_\alpha/2 \right) \right\rangle}\,.
\end{equation}
Without knowing the actual correlation between $\phi_\mathrm{pair, \alpha}$ and $\delta \phi_\alpha$, it is difficult to compare the size of $\Delta C_{123}/C_3\{2\}$ with the ratios in Eq.~(\ref{eq:Delta112_over_C2}). 

%
%=======================================
\begin{figure}[ht!]
  \includegraphics[width=0.95\linewidth]{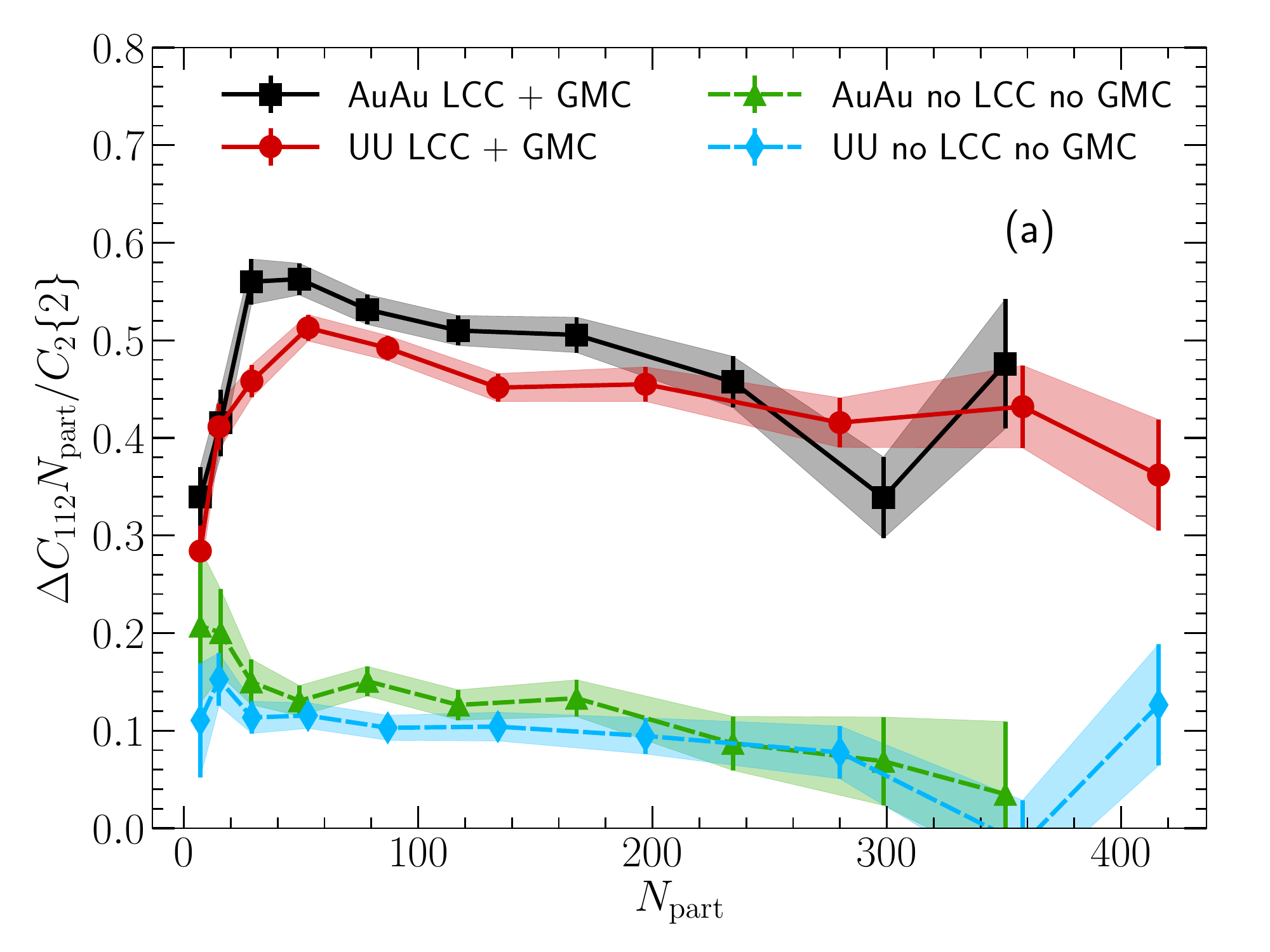}
  \includegraphics[width=0.95\linewidth]{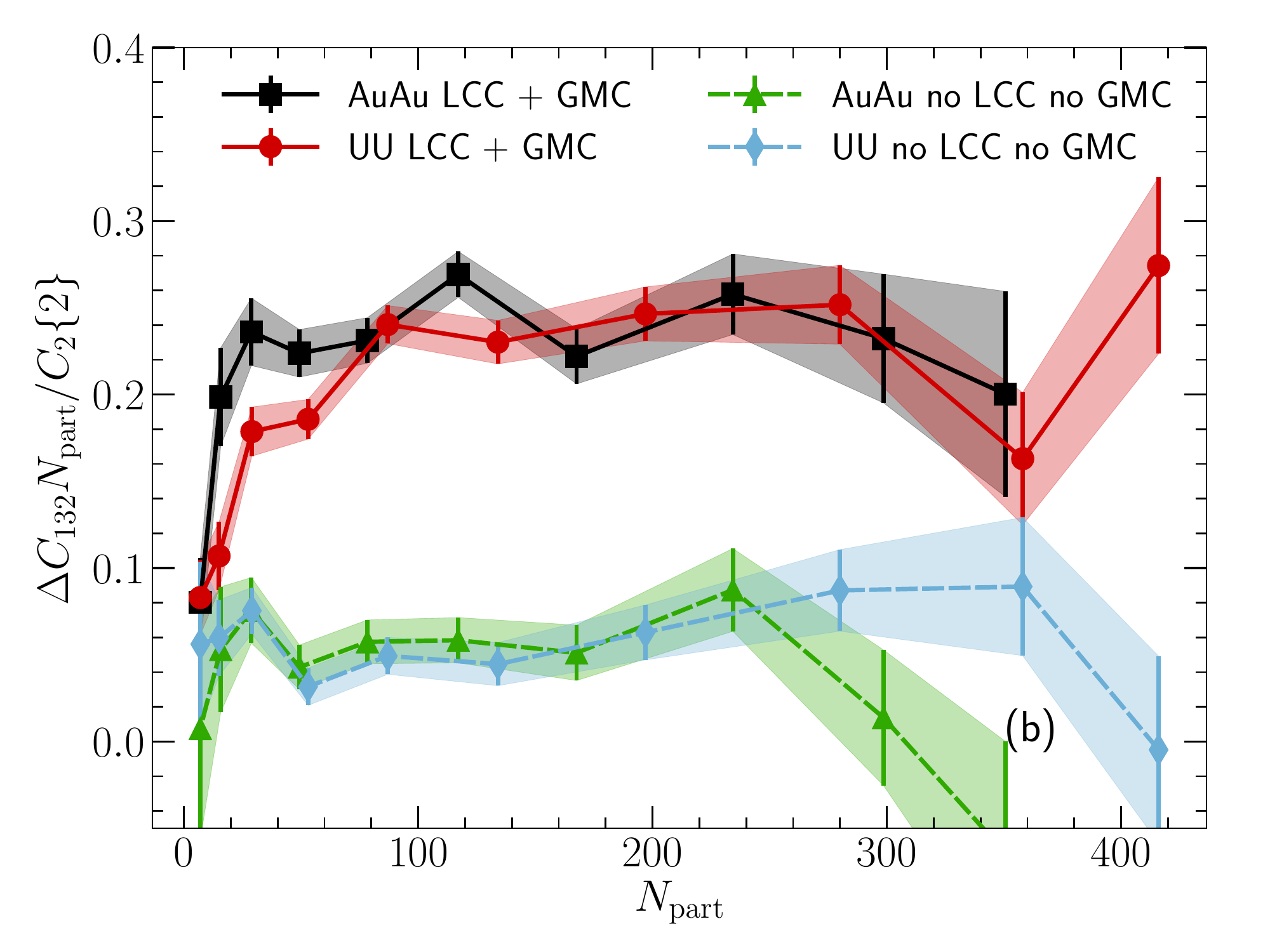}
  \includegraphics[width=0.95\linewidth]{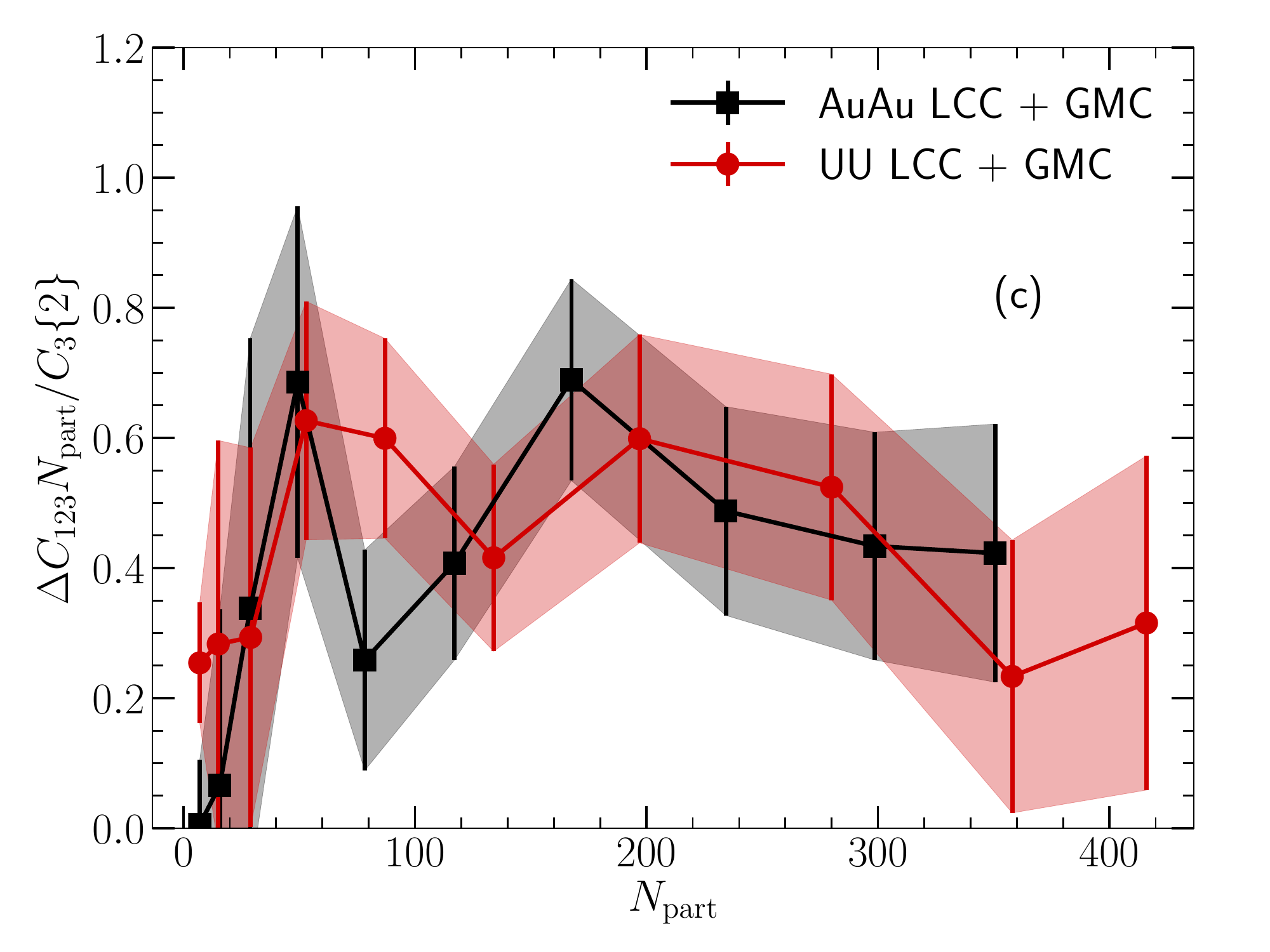}
  \caption{Panel (a): The difference between opposite-sign and same-sign three-particle correlation functions $\Delta C_{112} = C_{112}(\mathrm{OS}) - C_{112}(\mathrm{SS})$ is scaled by the value of $N_\mathrm{part}/C_2\{2\}$ in every centrality bin in Au+Au and U+U collisions. Panel (b) for $\Delta C_{132}N_\mathrm{part}/C_2\{2\}$ and panel (c) for $\Delta C_{123}N_\mathrm{part}/C_3\{2\}$. Here $C_2\{2\}$ and $C_3\{2\}$ are the second and the third order Fourier coefficients of the two-particle correlation of all charged particles.  
  Results with and without imposing local charge conservation (LCC) and global momentum conservation (GMC) are shown.}
  \label{fig10}
\end{figure}
%=======================================
%

Numerical results for $\Delta C_{mnk}$ are shown in Figs.\,\ref{fig10}. Including local charge and global momentum conservation, we find that $ \Delta C_{112}/C_2\{2\} \approx \Delta C_{123}/C_3\{2\}$ and $\Delta C_{112}/C_2\{2\} \approx 2 \Delta C_{132}/C_2\{2\}$. The three particle correlations $\Delta C_{mnk}$ scale reasonably well with $C_k\{2\}/N_\mathrm{part}$ from central to mid-peripheral centrality bins. The local charge conservation increases the absolute values of $\Delta C_{mnk}$ by a factor of five. There is no statistically significant difference in these scaled observables between Au+Au and U+U collisions. 

Based on Eqs.~(\ref{eq:DeltaC112}) and (\ref{eq:DeltaC132}), we can compute the following ratios to numerically test the factorization
\begin{eqnarray}
    \frac{N_\mathrm{ch}}{2} \frac{\Delta C_{112}}{C_2\{2\}} K_1 &=& \frac{\left\langle \cos(2(\phi_\mathrm{pair,\alpha} - \phi_c)) \right\rangle \left\langle \cos \left(\delta \phi_\alpha \right)\right\rangle}{\left\langle \cos(2(\phi_\mathrm{pair,\alpha} - \phi_c)) \cos \left(\delta \phi_\alpha \right) \right\rangle} \lesssim 1, \notag \\
    \frac{N_\mathrm{ch}}{2} \frac{\Delta C_{132}}{C_2\{2\}} \frac{K_1}{K_2} &=& \frac{\left\langle \cos(2(\phi_\mathrm{pair,\alpha} - \phi_c)) \cos \left(2\delta \phi_\alpha \right) \right\rangle \left\langle \cos \left(\delta \phi_\alpha \right)\right\rangle} {\left\langle \cos(2(\phi_\mathrm{pair,\alpha} - \phi_c)) \cos \left(\delta \phi_\alpha \right) \right\rangle \left\langle \cos \left(2\delta \phi_\alpha \right)\right\rangle} \notag \\
    && \sim 1\notag.
\end{eqnarray}
%
%=======================================
\begin{figure}[ht!]
  \includegraphics[width=0.95\linewidth]{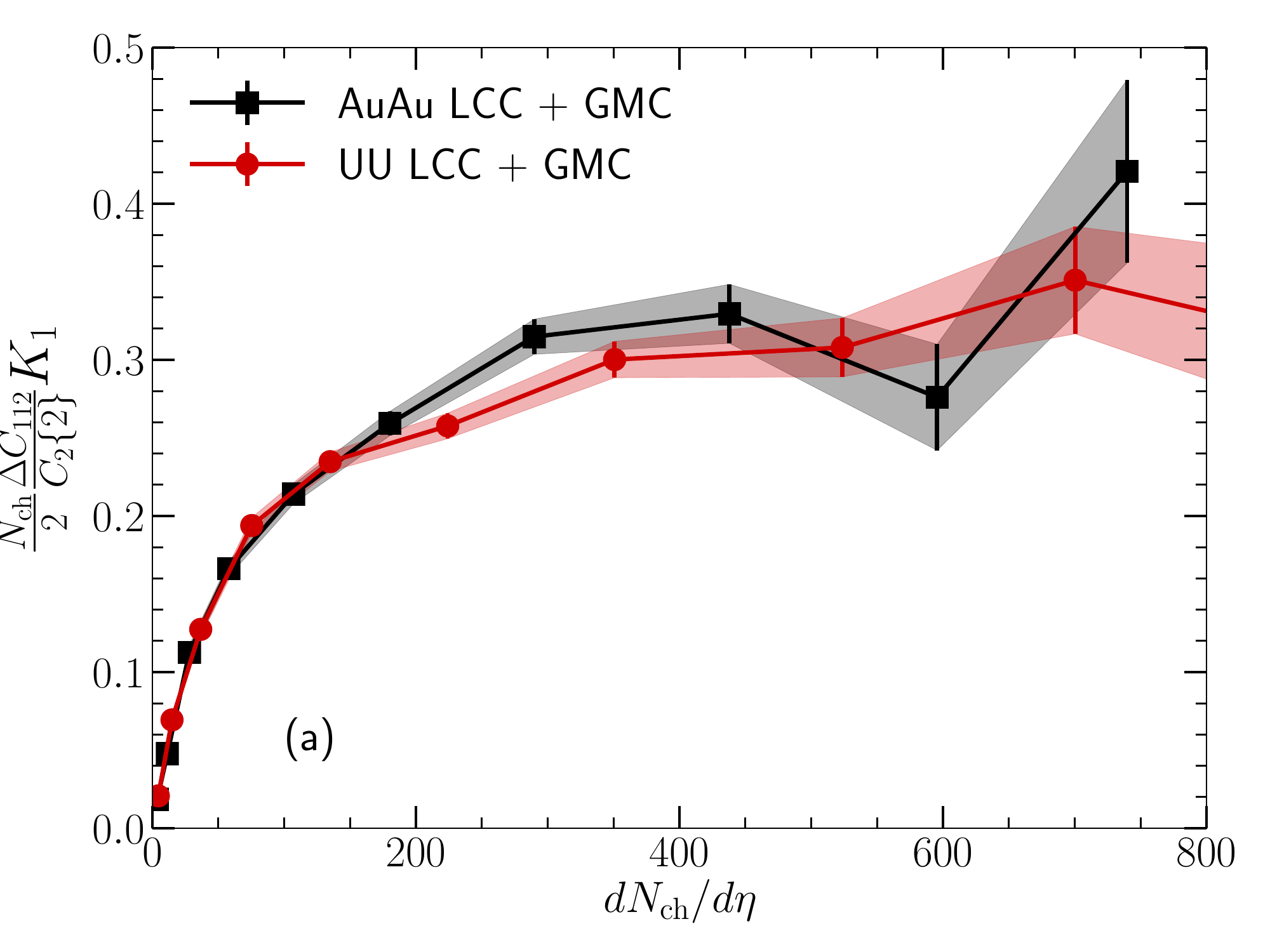}
  \includegraphics[width=0.95\linewidth]{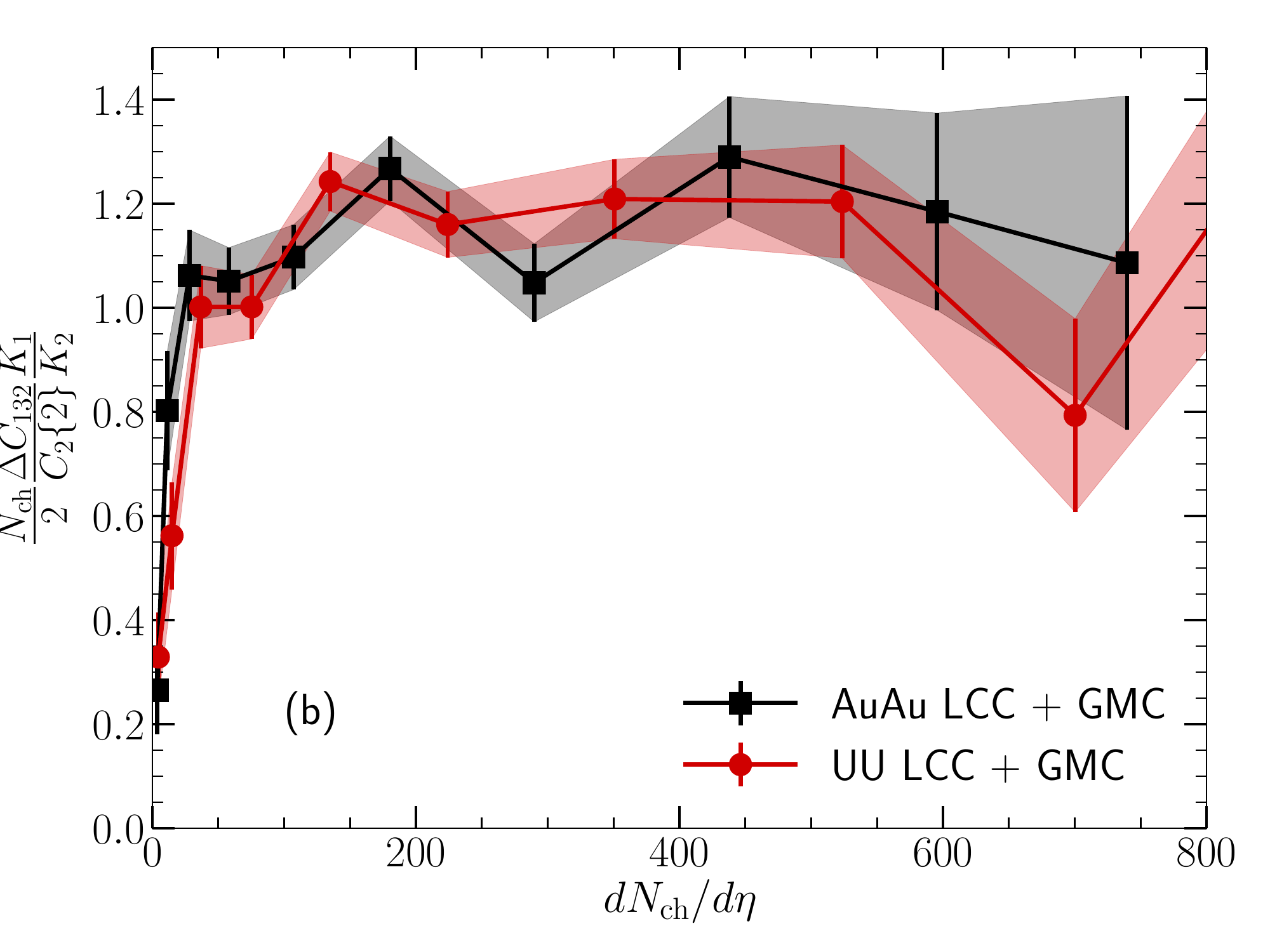}
  \caption{Panel (a): Test of the scaling of $\frac{N_\mathrm{ch}}{2} \frac{\Delta C_{112}}{C_2\{2\}} K_1$ as a function of centrality in Au+Au and U+U collisions. Panel (b) Test of the scaling of $\frac{\Delta C_{132}}{C_2\{2\}} \frac{K_1}{K_2}$.}
  \label{fig11}
\end{figure}
%=======================================
%
Those ratios are shown in Fig.~\ref{fig11}. The value of $\frac{N_\mathrm{ch}}{2} \frac{\Delta C_{112}}{C_2\{2\}} K_1$ is only $\sim 0.3$ suggesting that the factorization is badly broken. The situation is much better for scaling of $\Delta C_{132}$. This could be because of the cancellation of the correlations between the numerator and denominator. 

\begin{figure}[ht!]
  \includegraphics[width=1.0\linewidth]{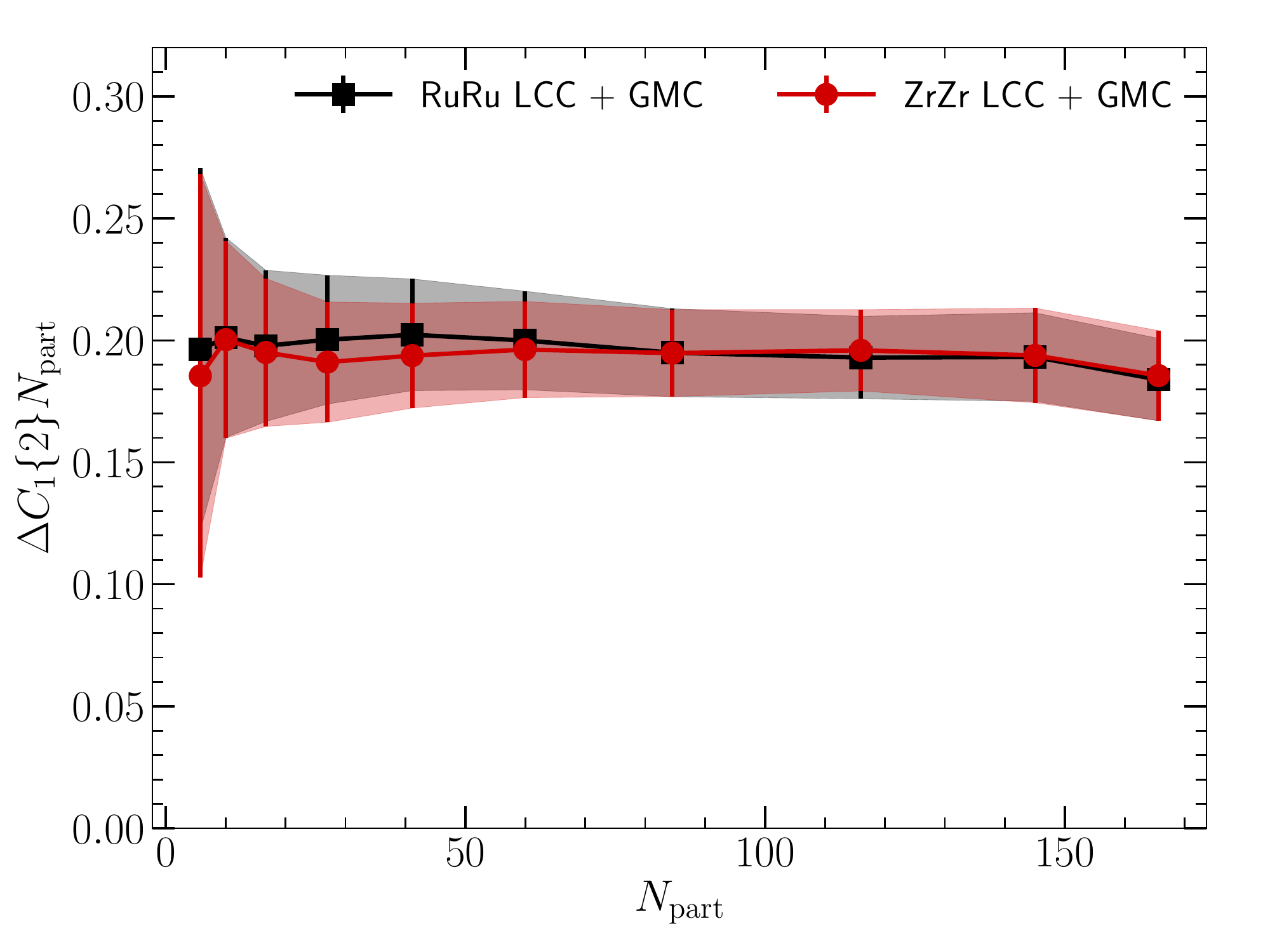}
  \caption{The difference between opposite-sign and same-sign two-particle correlation functions $\Delta C_1\{2\} = C_1\{2\}(\mathrm{OS}) - C_1\{2\}(\mathrm{SS})$ scaled by the number of participants $N_\mathrm{part}$ in Ru+Ru and Zr+Zr collisions. Results include local charge conservation (LCC) and global momentum conservation (GMC).}
  \label{fig13}
\end{figure}

\begin{figure}[ht!]
  \includegraphics[width=1.0\linewidth]{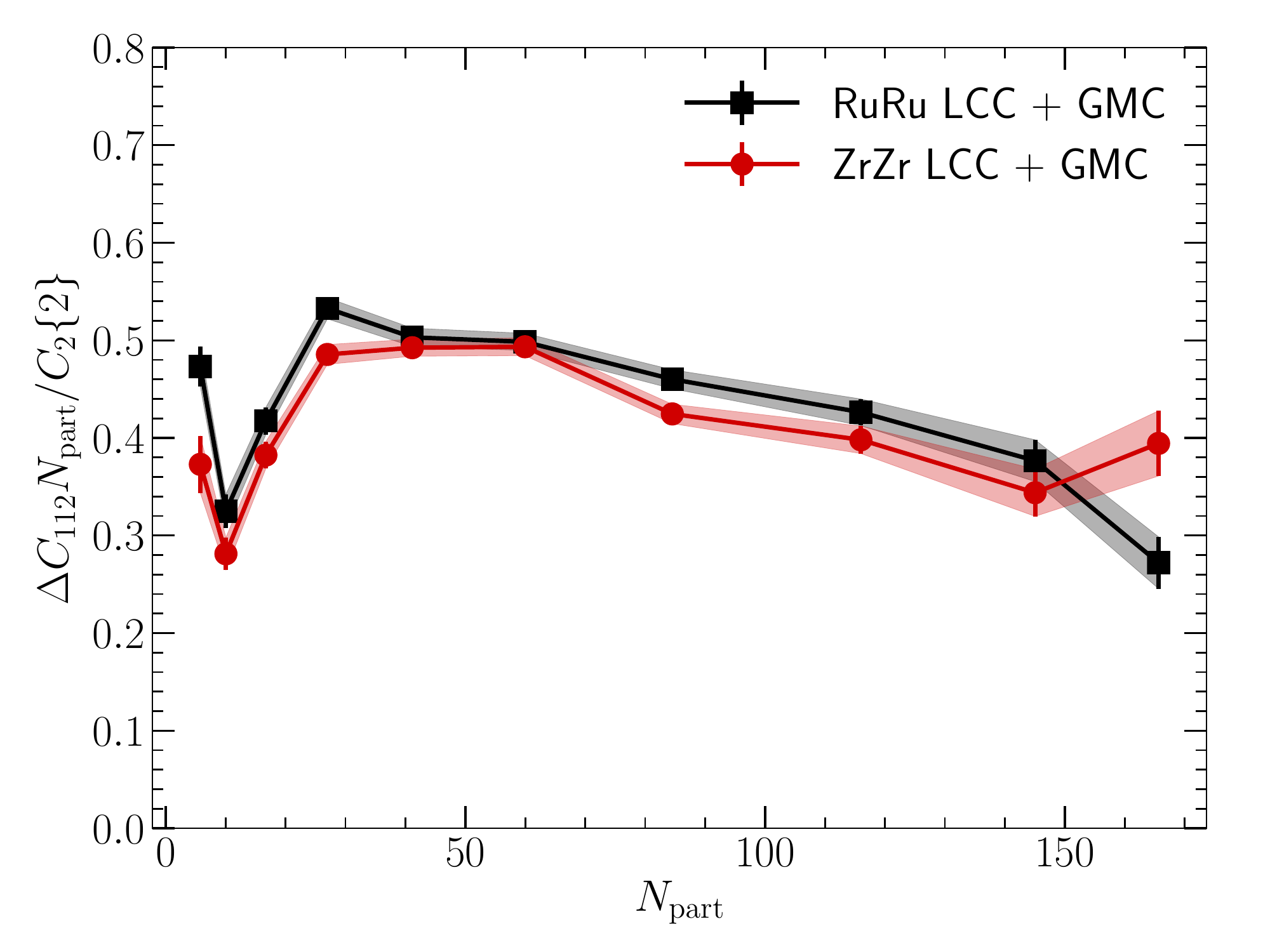}
  \caption{The difference between opposite-sign and same-sign three-particle correlation functions $\Delta C_{112} = C_{112}(\mathrm{OS}) - C_{112}(\mathrm{SS})$ is scaled by the value of $N_\mathrm{part}/C_2\{2\}$ in every centrality bin in Ru+Ru and Zr+Zr collisions. Here $C_2\{2\}$ is the second Fourier coefficient of the two-particle correlation of all charged particles. Results include local charge conservation (LCC) and global momentum conservation (GMC).}
  \label{fig14}
\end{figure}

Predictions of charge dependent multi-particle correlations in isobar collisions (Ru+Ru and Zr+Zr) are shown in Figs.~\ref{fig13} and \ref{fig14}. The two particle correlation functions $\Delta C_1\{2\}$ in Ru+Ru and Zr+Zu collisions scale very well with the number of participants $N_\mathrm{part}$. The values of $\Delta C_1\{2\}N_\mathrm{part}$ are very close to those shown in Au+Au and U+U collisions shown in Fig.~\ref{fig9}. The three particle correlations $\Delta C_{112}/C_2\{2\}$ are approximately the same in Ru+Ru and Zr+Zr collisions. This is because the hydrodynamic flow backgrounds in these two collision systems are very close to each other as shown in Fig.~\ref{fig5}. Our results provide a realistic background baseline for the search of the Chiral Magnetic Effect in upcoming RHIC isobar data. The difference between Ru+Ru and Zr+Zr visible in Fig.\,\ref{fig14} is small in comparison to the expected $\sim 10-15\%$ difference generated by the CME \cite{Shi:2017cpu,Deng:2018dut,Sun:2018idn}.

We note that our results for charge dependent correlators calculated including local charge conservation should be taken as upper bounds. Although we anticipate a direct comparison of RHIC measurements with our predictions, any conclusions from such data-model comparison should account for the approximations in our approach. This is because the simple implementation of producing a negatively charged particle for every positively charged particle in the same freeze-out surface element leads to a maximal correlation between opposite sign charges. In the future, it will be interesting to investigate more realistic implementations of local charge conservation along with prescriptions for local momentum conservation in the particle sampler.
%
%=======================================
%=======================================
%
%
%=======================================
%=======================================
%

\section{Conclusions} \label{sec:conc}
We have presented results for charge inclusive and charge dependent multi-particle correlation functions in heavy ion collisions at top RHIC energy from calculations in a hybrid framework based on the IP-Glasma initial state,  \textsc{Music} viscous fluid dynamics simulations, and the UrQMD hadronic cascade. In the sampling of particles, required to transition from the fluid to the microscopic transport regime, we have implemented explicit prescriptions to impose  global momentum and local charge conservation. 

After adjusting the free parameters, such as shear and bulk viscosities to achieve a good description of particle multiplicities, mean transverse momentum and anisotropic flow, we make predictions for charge inclusive three and four particle correlators, as well as two and three particle charge dependent correlators.

The studied charge inclusive correlators provide a measure of correlations between flow harmonics of different order. We studied three and four particle correlations, in particular symmetric cumulants and mixed harmonic event-plane correlators. We found good agreement with the experimental data in Au+Au collisions at RHIC, except for $C_{112}$, which is expected to be very sensitive to local momentum conservation in the particle sampling procedure, which is not yet included in our framework. The observed good agreement of our calculation with the majority of these new multi-particle observables from RHIC further validates our hybrid framework. We note that the hadronic afterburner plays an important role for describing this data at RHIC (c.f. \cite{Adamczyk:2017byf}, where the effect of the afterburner is demonstrated).

As expected, charge dependent multi-particle correlators were shown to be very sensitive to whether local charge conservation in the process of particle sampling was implemented. Besides presenting numerical results, we analyzed the structure of the charge dependent two and three particle correlators for our implementation of LCC, and explored scaling relations of the correlators with particle multiplicity and provided expectations in case of certain factorization conditions. We further expressed the charge dependent three particle correlators in terms of two particle correlations and demonstrated where these relations hold in the full hybrid framework calculation.

Our analysis of charge dependent correlators provides an important estimate of the background for a potential chiral magnetic effect signal in heavy ion collisions. In particular, our predictions for the isobar collision systems Ru+Ru and Zr+Zr will be very important in this context. For $\Delta C_{112}$, the main observable sensitive to the CME, we predict small differences between Ru+Ru and Zr+Zr collisions, caused by the difference in the shapes of the Ru and Zr nuclei. We note that the lack of local momentum conservation in the particle sampling implementation, which leads to disagreement with the experimentally observed charge inclusive $C_{112}$ correlator, should have little effect on the difference between opposite sign and same sign expressions. Thus, our calculations should provide a solid prediction for $\Delta C_{112}$ in the absence of the CME.

In the future, we will study the rapidity dependence of the multi-particle correlation functions. It requires to include longitudinal fluctuations in the initial state model, as discussed e.g. in Refs.~\cite{Shen:2017bsr,Schenke:2016ksl}. Further coupling of our simulations with the evolution of electromagnetic fields \cite{Gursoy:2018yai} will allow the direct study of CME signals on top of the event-by-event flow background.

\section*{Acknowledgments}
BPS and PT are supported under DOE Contract No. DE-SC0012704. CS is supported under DOE Contract No. DE-SC0013460. This research used resources of the National Energy Research Scientific Computing Center, which is supported by the Office of Science of the U.S. Department of Energy under Contract No. DE-AC02-05CH11231. This work is supported in part by the U.S. Department of Energy, Office of Science, Office of Nuclear Physics, within the framework of the Beam Energy Scan Theory (BEST) Topical Collaboration.

\appendix

\section{Effects of precise matching of the Yang-Mills and hydrodynamic $T^{\mu\nu}$} \label{app:A}
In this appendix we study the effect of including the initial viscous part of the energy momentum tensor of the Yang-Mills fields provided by the IP-Glasma model, as well as the effect of smooth matching of the pressure. Both contributions have not been included in previous calculations (e.g. \cite{Gale:2012rq,Ryu:2015vwa,McDonald:2016vlt}) because an effective completely efficient thermalization process was assumed to occur before the hydrodynamic simulations start.

It seems more and more unlikely that thermalization will happen to such a degree, and arguments that viscous hydrodynamics should be applicable in systems relatively far from complete local equilibrium (for a review see \cite{Florkowski:2017olj}) motivate us to include these non-equilibrium contributions at the initial time of the hydrodynamic simulation.

The shear viscous correction will generate a longitudinal pressure of approximately zero and include spatially dependent transverse terms. The Yang-Mills calculation is conformal and has no bulk viscosity, but we use the freedom of having an initial $\Pi$ to smoothly match the pressure in the Yang-Mills phase to the pressure of the lattice QCD EoS. Its sign is opposite to the typical bulk viscous correction and thus generates an additional outward push.

We studied the effect of these contributions by turning them off sequentially in 20-30\% central Au+Au collisions and analyzing the change in $\langle p_T \rangle$,  $v_2\{2\}$, and $v_3\{2\}$. Results are presented in Table \ref{tab:initialPi}.

We find that both including the initial shear stress as well as the smooth matching of the pressure have a non-negligible effect on these observables, in particular the $v_n$. All three quantities increase with the inclusion of the initial $\pi^{\mu\nu}$ and the effective $\Pi = \varepsilon/3-P_{\rm lat}$, where $P_{\rm lat}$ is the pressure from the lattice QCD EoS. For $v_3$ the effect is largest with a change greater than $50\%$.

\begin{table}[hb]
\begin{center}  
    \begin{tabular}{ | c | c | c | c |}
    \hline
          & $\langle p_T \rangle$ (GeV) & $v_2\{2\}$ & $v_3\{2\}$ \\ \hline
    $T_\mathrm{ideal}^{\mu \nu}$ & $0.5071(5)$  & $0.047(1)$  & $0.0125(4)$  \\ \hline
    $T_\mathrm{ideal}^{\mu \nu} + \pi^{\mu\nu}$ & $0.5236(6)$ & $0.056(1)$  & $0.0138(4)$  \\ \hline
    full $T^{\mu \nu}$ & $0.5546(2)$  & $0.066(1)$ & $0.0194(3)$ \\
    \hline
    \end{tabular}
\end{center}
\caption{The effects of initial shear stress tensor and initial bulk pressure on charged hadron mean-$p_T$ and $v_{2,3}\{2\}$ flow coefficients. The numbers in parentheses show the statistical errors on the last digit. \label{tab:initialPi}}
\end{table}

\bibliography{ref}

\end{document}